\documentclass[12pt]{article}
\usepackage[centertags]{amsmath}
\usepackage{amsfonts}
\usepackage{amssymb}
\usepackage{amsthm}
\usepackage{newlfont}
\usepackage{hyperref}

\numberwithin{equation}{section}

\newtheorem{theorem}{Theorem}

\newtheorem{definition}{Definition}

\begin{document}

\title{A revision for Heisenberg uncertainty relation based on environment variable in the QCPB theory}

\author{ Gen  WANG\thanks{School of Mathematical Sciences, Xiamen University,
     Xiamen, 361005, P.R.China. email:
wanggen@zjnu.edu.cn }}

\date{ }

\maketitle

\begin{abstract}
The Heisenberg uncertainty principle and its extensions are all still inequalities form which hold the superior approximate estimations.
Based on quantum covariant Poisson bracket theory, we propose quantum geomertainty relation to modify and explain the uncertainty relation to positively give a complete description of reality that enhances the outcome of each measurement with certainty.  It demonstrates that entanglement term exists between the observable and the environment and nicely explains how the environment has an effect on the measurement which causes the unavoidable influences.

\end{abstract}

\tableofcontents

\section{Introduction}
 In March 1926, Heisenberg realized that the non-commutativity implies the uncertainty principle. This implication provided a clear physical interpretation for the non-commutativity.  Heisenberg showed that the QPB defined by commutation relation implies an uncertainty.  Any two variables that do not commute cannot be measured simultaneously--the more precisely one is known, the less precisely the other can be known.  Later, the uncertainty principle historically introduced first in 1927, by Heisenberg \cite{1}, it states that the more precisely the position of some particle is determined, the less precisely its momentum can be known, and vice versa.

The Heisenberg uncertainty relation for canonical observables $q$ and $p$ is certainly one of the most fundamental results in quantum mechanics. It was introduced by Heisenberg and mathematically proved by Kennard \cite{2} and Weyl \cite{3}. At present, the Heisenberg uncertainty relation has richly developed. Later on the Heisenberg uncertainty relation was generalized to the case of two arbitrary observables by Robertson \cite{4,5} and Schr\"{o}dinger \cite{6}. In fact in \cite{5,6} an improved version of the Heisenberg uncertainty relation has been obtained. Finally, Robertson \cite{7} was able to extend the previous results to an arbitrary number of observables. The inequalities found in \cite{7} are called the Heisenberg-Robertson and Schr\"{o}dinger uncertainty relations. Recently a great deal of interest in uncertainty relations is observed. It has been shown that they can be used to define squeezed and coherent states and also to generalize this important concepts by introducing the notion of intelligent states \cite{8,9,10,11,12,13,14}.
It seems to be natural that any theory which would like to describe quantum systems should reproduce in some sense the uncertainty relations. So we expect that it must be also the case in deformation quantization\cite{17,18,19,20,21,22}.  Nowadays, it has already been put into a stronger form and other versions, even generalized uncertainty relation in string theory. In a recent paper  \cite{15} that has well proposed the QCPB equipped with quantum geometric bracket to define quantum covariant Hamiltonian system (QCHS) and covariant dynamics,  generalized Heisenberg equation, G-dyamics, which have given a series of concise and complete interpretation of the quantum mechanics.

Heisenberg used uncertainty principle to explain how the measurement would destroy the classical characteristics of quantum mechanics. The well-known double slit interference is an example. However, some quantum physicists believe that it is not Heisenberg's uncertainty principle that can explain the disappearance of interference patterns, but some other mechanisms.
There are still many problems around the uncertainty principle that have not been well explained.

\subsection{The uncertainty relation}
The uncertainty principle is one of the most characteristic and important consequences of the new quantum mechanics. This principle, as formulated by Heisenberg for two conjugate quantum-mechanical variables,
states that the accuracy with which two such variables can be measured simultaneously is subject to the restriction that the product of
the uncertainties in the two measurements is at least of order $\hbar/2$ \cite{23,24,25,26}.  Heisenberg's uncertainty principle is represented by the statement that the operators corresponding to certain observables do not commute.

The uncertainty principle implies that it is in general not possible to predict the value of a quantity with arbitrary certainty, even if all initial conditions are specified.  In order to explain the physical reasons for the uncertainty relation related to the quantum measurement, scientists have been thinking about how the environment acts on the measurement, for this reason, the physicists have presented so many theories to explain such phenomenon, but they still face some unsolved problems in this.

Why not be unable to certainly measure? why can't predict accurately? What's the reason for this uncertainty?  The most frequently reason mentioned is the interference of measurement from the environment. The only consensus we have made is the uncertainties in measurements caused by the environment, the deep reason for this still seeks.

In this section, it requires some knowledge of Heisenberg uncertainty relation and its related extension.  Firstly, we briefly give basic quantum knowledge,   in retrospect.
For an arbitrary Hermitian operator,  we can associate a standard deviation
\[{{\sigma }_{F}}=\sqrt{\langle {{\hat{F}}^{2}}\rangle -{{\langle \hat{F}\rangle }^{2}}}\]
where the brackets $ \langle \hat{F}\rangle $ indicate an expectation value, sometime, we also use $\overline{\hat{F}}$ to express an expectation value.
More precisely, for a given operator $\hat{F}$ and its expectation value $\overline{F}$, we have
\begin{align}
\overline{{{\left( \hat{F}-\overline{F} \right)}^{2}}}  &=\overline{{{\hat{F}}^{2}}-2\hat{F}\overline{F}+{{\overline{F}}^{2}}}=\overline{{{F}^{2}}}-2\overline{\hat{F}}\overline{F}+{{\overline{F}}^{2}} \notag\\
 & =\overline{{{F}^{2}}}-2{{\overline{F}}^{2}}+{{\overline{F}}^{2}}
 =\overline{{{F}^{2}}}-{{\overline{F}}^{2}}\notag
\end{align}
where $\overline{\hat{F}}=\overline{F}$ has been used, we will use $\overline{\hat{F}}=\overline{F}=\left\langle \hat{F} \right\rangle,$  and
${{\sigma }_{A}}=\Delta A={{\sigma }_{\hat{A}}}$.

The mean error of mechanical quantity operator $\hat{A}$ and wave function $\psi$, is defined $\Delta \hat{A}=\hat{A}-\left( \psi ,\hat{A}\psi  \right)=\hat{A}-\overline{A}$ and then a series of operation accordingly follow, such as the mean square error of mechanical quantity $\hat{A}$ is
\begin{align}
 {{\left( \Delta A \right)}^{2}} &\equiv \overline{{{\left( \Delta \hat{A}\right)}^{2}}}=\left( \psi ,{{\left( \Delta\hat{A} \right)}^{2}}\psi  \right) \notag\\
 & =\left( \psi ,{{\left( \hat{A}-\overline{A} \right)}^{2}}\psi  \right)=\left( \psi ,\left( {{\hat{A}}^{2}}-2\hat{A}\overline{A}+{{\left( \overline{A} \right)}^{2}} \right)\psi  \right) \notag\\
 & =\left( \psi ,{{\hat{A}}^{2}}\psi  \right)-{{\left( \overline{A} \right)}^{2}}=\overline{{{A}^{2}}}-{{\left( \overline{A} \right)}^{2}} \notag
\end{align}The eigenvalues equation is $\hat{A}\psi =A\psi$, and expectation value follows $\overline{A}=\left( \psi ,\hat{A}\psi  \right)$. The square root of it is\[\Delta A=\sqrt{\overline{{{\left( \Delta \hat{A} \right)}^{2}}}}=\sqrt{\overline{{{A}^{2}}}-{{\left( \overline{A} \right)}^{2}}}=\sqrt{\left( \psi ,{{\hat{A}}^{2}}\psi  \right)-{{\left( \psi ,\hat{A}\psi  \right)}^{2}}}\]
Heisenberg's uncertainty principle shows that some physical quantities describing a micro particle, such as position and momentum, momentum moment, time and energy, can not have a definite value at the same time. The more definite one quantity is, the more uncertain the other quantity is. That is, the inequality satisfied by the fluctuation of any two mechanical quantities in an arbitrary quantum state.

Let any two mechanical quantities $\hat{F}$ and $\hat{G}$ be given, then $\hat{F}$ and $\hat{G}$ are Hermitian operator, Heisenberg's relation is shown as
\begin{equation}\label{eq1}
  \sqrt{\overline{{{F}^{2}}}\cdot \overline{{{G}^{2}}}}\ge \frac{\left| \overline{\left[ \hat{F},\hat{G} \right]_{QPB}} \right|}{2}
\end{equation}If commutation relation between $\hat{F},\hat{G}$ is $\left[ \hat{F},\hat{G} \right]_{QPB}=\sqrt{-1}\kappa $, then
above inequality can write in the form $\Delta F\cdot \Delta G\ge \frac{\kappa }{2}$, above inequality holds for any two Hermitian operators $\hat{F}$ and $\hat{G}$, $\overline{F}$ and $\overline{G}$ are real numbers.
Any two mechanical quantities $\hat{F}$ and $\hat{G}$ should satisfy the relation under the fluctuation of any quantum state, that is, uncertainty relation.

The commutation of the position and momentum comply with \cite{26}
\begin{equation}
  \left[ {{x}_{i}},{{\hat{p}}_{j}} \right]_{QPB}=\sqrt{-1}\hbar {{\delta }_{ij}}
\end{equation}Where
${{\delta }_{ij}}=\left\{ \begin{matrix}
   1,i =j  \\
   0,i \ne j   \\
\end{matrix} \right.$.
The formal inequality relating the standard deviation of position $\sigma_{x}$ and the standard deviation of momentum $\sigma_{p}$ was derived by Kennard,  later that year and by Hermann Weyl in 1928:
$\sigma _{x}\sigma _{p}\geq {\frac {\hbar }{2}}$,
where $\hbar$ is the reduced Planck constant.

Some concepts are given such as eigenfunction, eigenvalues and mean value, more precisely, the eigenvalues equation is $\hat{A}\varphi =A\varphi $, where $A$ is eigenvalue, and $\varphi$ is eigenfunction, the mean values for any observables can be expressed as
\[\left\langle A \right\rangle =\overline{A}=\left\langle  \varphi  \right|\hat{A}\left| \varphi  \right\rangle \]
In matrix mechanics, observables such as position and momentum are represented by self-adjoint operators. When considering pairs of observables, an important quantity is the commutator. For a pair of operators $\hat {a}$ and $\hat {b}$, one defines their commutator as
$[{\hat {a}},{\hat {b}}]_{QPB}={\hat {a}}{\hat {b}}-{\hat {b}}{\hat {a}}$.
In the case of position and momentum, the commutator is the canonical commutation relation
$[{\hat {x}},{\hat {p}}]_{QPB}=\sqrt{-1}\hbar$.
The physical meaning of the non-commutativity can be understood by considering the effect of the commutator on position and momentum eigenstates.
\[\Delta x\cdot \Delta p-\left| \frac{\overline{{{\left[ x,\hat{p} \right]}_{QPB}}}}{2\sqrt{-1}} \right|\ge 0\]
For example, if a particle's position is measured, then the state amounts to a position eigenstate. This means that the state is not a momentum eigenstate, however, but rather it can be represented as a sum of multiple momentum basis eigenstates. In other words, the momentum must be less precise. This precision may be quantified by the standard deviations,
$${\displaystyle \sigma _{x}={\sqrt {\langle {\hat {x}}^{2}\rangle -\langle {\hat {x}}\rangle ^{2}}}},~~
{\displaystyle \sigma _{p}={\sqrt {\langle {\hat {p}}^{2}\rangle -\langle {\hat {p}}\rangle ^{2}}}.}$$
As in the wave mechanics interpretation above, one sees a tradeoff between the respective precisions of the two, quantified by the uncertainty principle.

\subsection{Generalized uncertainty principle}
Heisenberg uncertainty principle is the cornerstone of quantum mechanics. It elaborates the relationship between the degree of uncertainty between two related variables. At present, many different versions of the expression are obtained around the relationship of uncertainty.  The research on uncertain relationship is still very active. GUP is the generalized uncertainty principle \cite{28,29,30}, which modifies the common uncertainty principle.
(Witten 1996; Amati et al .1987; Yoneya 2000; Adler 1999; Maggiore 1994 ; Veneziano 1980; Farmany and Dehghani 2010).  It discusses and considers that the principle of uncertainty should take form
\begin{equation}\label{eq11}
  \Delta x>\frac{\hbar }{\Delta p}+\alpha l_{p}^{2}\frac{\Delta p}{\hbar }
\end{equation}
where $\alpha$ is a dimensionless constant,
and $ {{l}_{p}}={{\left( G\hbar /{{c}^{3}} \right)}^{1/2}}\approx 1.6\times {{10}^{-35}}m$ is Planck length. Normally set to the order of unity but
which in string theory is found to correspond to the string tension. The second term in \eqref{eq11} relates to the uncertainties due to the gravitational effects, and so they will only become significant when $\Delta x\approx {{l}_{p}}$.

Along this line, many studies have converged on the idea that the HUP should be properly modified at the quantum gravity scale, in order to accommodate the
existence of such a fundamental length. In this sense, one of the most adopted generalizations of the uncertainty principle (GUP) reads \cite{30}
\begin{equation}\label{eq12}
  \delta x \delta p \geq \frac{\hbar}{2} \pm 2|\beta| \ell_{p}^{2} \frac{\delta p^{2}}{\hbar}=\frac{\hbar}{2} \pm 2|\beta| \hbar \frac{\delta p^{2}}{m_{p}^{2}}
\end{equation}
where the sign $\pm$ refers to positive/negative values of
the dimensionless deformation parameter $\beta$, which is assumed to be of order unity in some models of quantum gravity, and in particular in string theory.
\eqref{eq12} can be deduced from the modified commutator
$$[\hat{x}, \hat{p}]_{QPB}=\sqrt{-1} \hbar\left(1 \pm|\beta|\left(\frac{\hat{p}^{2}}{m_{p}^{2}}\right)\right)$$
One of the most difficult and stimulating challenges the
physics community has been struggling with for a long
time is to understand whether the gravitational interaction has an intrinsic quantum nature and, if so, how to formulate a thorough quantum theory of gravity which avoids conceptual problems and is able to make successful predictions at any energy scale. In pursuing the aim of combining both gravitational and quantum effects, the question inevitably arises as to whether the basic principles of quantum mechanics need to be revised in the
quantum gravity realm \cite{30}.

An important feature of the existence of a minimal length is the modification of the standard Heisenberg commutation relation in the usual quantum mechanics. Such relations are known as the Generalized Uncertainty Principle (GUP). In one dimension, the simplest form of such relations can be written as \cite{28}
$$\triangle p \triangle x \geq \frac{\hbar}{2}\left(1+\beta(\triangle p)^{2}+\gamma\right)$$
where $\beta$ and $\gamma$ are positive and independent of $\triangle x$  and $\triangle p$, but may in general depend on the expectation values $\left\langle x \right\rangle $ and $\left\langle p \right\rangle$.
 The usual Heisenberg commutation relation can be recovered in the
limit $\beta=\gamma= 0$.  As is clear from above equation, this equation implies a minimum position uncertainty of $(\triangle x)_{\min }=\hbar \sqrt{\beta}$, and hence $\beta$ must be related to the Planck length. For a more general discussion on such deformed Heisenberg algebras, especially in three dimensions. Now, it is possible to realize above equation from the following commutation relation between position and momentum operators \cite{28}
$$[\hat{x}, \hat{p}]_{QPB}=\sqrt{-1}\hbar\left(1+\beta p^{2}\right)$$
where we take $\gamma=\beta\langle p\rangle^{2}$.
More general cases of such commutation relations are studied.
Also various applications of the low energy effects of the modified Heisenberg uncertainty relations have been extensively studied.

The general Weyl-Heisenberg algebra of the kind studied in \cite{49,50,51}
$$\left[x_{i}, p_{j}\right]_{QPB}=\sqrt{-1} \hbar\left(\delta_{i j}\left(1+\alpha_{1} p+\beta_{1} p^{2}\right)+p_{i} p_{j}\left(\frac{\alpha_{2}}{p}+\beta_{2}\right)\right)$$
where $p\equiv \sqrt{{{p}^{2}}}=\sqrt{{{p}_{i}}{{p}^{i}}}$ , and $\alpha $ and $\beta $ must be dimensionfull constants, $\left[ {{\alpha }_{a}} \right]\sim 1/p,\left[ {{\beta }_{a}} \right]\sim 1/{{p}^{2}}$, for $a = 1,2$.

\subsection{Other generalized form of uncertainty relation}
Heisenberg's uncertainty principle is one of the main tenets of quantum theory. Nevertheless, and despite its fundamental importance for our understanding of quantum foundations, there has been some confusion in its interpretation.  Although Heisenberg's first argument was that the measurement of one observable on a quantum state necessarily disturbs another incompatible observable, standard uncertainty relations typically bound the indeterminacy of the outcomes when either one or the other observable is measured.  If two incompatible observables cannot be measured together, one can still approximate their joint measurement, at the price of introducing some errors with respect to the ideal measurement of each of them \cite{32,33,34,35,36,37,38,39,40,41,42,43,44,45,46,47,48}.

For operators $\hat{A}$ and $\hat{B}$, the commutator for them is $\left[ \hat{A},\hat{B} \right]_{QPB}=\hat{A}\hat{B}-\hat{B}\hat{A}$, in this expression, the most common general form of the uncertainty principle is the Robertson uncertainty relation \cite{2,4} that reads
$$ {{\sigma }(A)}{{\sigma }(B)}\ge \left| \frac{1}{2\sqrt{-1}}\langle [\hat{A},\hat{B}]_{QPB}\rangle  \right|=\frac{1}{2}\left| \langle [\hat{A},\hat{B}]_{QPB}\rangle  \right|$$
where ${{\sigma }(A)}$ and ${{\sigma }(B)}$ are respectively standard variance of $A$ and $B$. In this notation, the Heisenberg-Robertson uncertainty relation in terms of $x$ and $\hat{p}$ is shown as
$$\vartheta \left( x\right)\vartheta \left( \hat{p} \right)\ge \frac{1}{2}\left| \langle \psi |\left[ x,\hat{p}\right]_{QPB}\left| \psi  \right\rangle  \right|$$
where $\vartheta $ can be $\Delta $ or $\sigma $.

Definitely, the QPB is an incomplete description, hence, it should be replaced by the complete QCPB theory, more precisely, ${{\left[ \hat{A},\hat{B}\right]}_{QPB}}\to \left[ \hat{A},\hat{B} \right]$.

The Robertson uncertainty relation immediately follows from a slightly stronger inequality---the Schr\"{o}dinger uncertainty relation that is given by
 \[{{\sigma }_{A}}{{\sigma }_{B}}\ge \sqrt{{{\left( \frac{1}{2}\langle \{\hat{A},\hat{B}\}_{ir}\rangle -\langle \hat{A}\rangle \langle \hat{B}\rangle  \right)}^{2}}+{{\left( \frac{1}{2\sqrt{-1}}\langle [\hat{A},\hat{B}]_{QPB}\rangle  \right)}^{2}}}\]
where we have introduced the anticommutator  $\{\hat{A},\hat{B}\}_{ir}=\hat{A}\hat{B}+\hat{B}\hat{A}$.
\begin{align}
  & \sigma _{A}^{2}=\langle (\hat{A}-\langle \hat{A}\rangle )\psi |(\hat{A}-\langle \hat{A}\rangle )\psi \rangle =\langle f|f\rangle ,\notag \\
 & \sigma _{B}^{2}=\langle (\hat{B}-\langle \hat{B}\rangle )\psi |(\hat{B}-\langle \hat{B}\rangle )\psi \rangle =\langle g|g\rangle \notag \\
 & \sigma _{A}^{2}\sigma _{B}^{2}=\langle f|f\rangle \langle g|g\rangle\geq{{\left| \left\langle  f \right|\left. g \right\rangle  \right|}^{2}}\notag
\end{align}
Since $ \langle f|g\rangle $ is in general a complex number, modulus squared of any complex number $z$ is defined as $|z{{|}^{2}}=z{{z}^{*}}$, so according to the complex variables function theory ,it has the following
\begin{align}
& \langle f|g\rangle =\langle (\hat{A}-\langle \hat{A}\rangle )\psi |(\hat{B}-\langle \hat{B}\rangle )\psi \rangle , \notag\\
 & |\langle f|g\rangle {{|}^{2}}={{\left(\frac{\langle f|g\rangle +\langle g|f\rangle }{2}\right)}^{2}}+{{\left(\frac{\langle f|g\rangle -\langle g|f\rangle }{2\sqrt{-1}}\right)}^{2}}\notag
\end{align}
where using the fact that $\hat{A}$ and $\hat{B}$ are Hermitian operators,
\begin{align}
  & \langle f|g\rangle -\langle g|f\rangle  =\langle [\hat{A},\hat{B}]_{QPB}\rangle  ,~~~ \langle f|g\rangle +\langle g|f\rangle =\langle \{\hat{A},\hat{B}\}_{ir}\rangle -2\langle \hat{A}\rangle \langle \hat{B}\rangle  \notag\\
 & |\langle f|g\rangle {{|}^{2}}={{\left(\frac{1}{2}\langle \{\hat{A},\hat{B}\}_{ir}\rangle -\langle \hat{A}\rangle \langle \hat{B}\rangle \right)}^{2}}+{{\left(\frac{1}{2\sqrt{-1}}\langle [\hat{A},\hat{B}]_{QPB}\rangle\right)}^{2}}. \notag
\end{align}
Therefore, in this way, the Schr\"{o}dinger uncertainty relation is given by above form.
This proof has an issue related to the domains of the operators involved. For the proof to make sense, the vector $\hat{B}\left| \psi  \right\rangle $  has to be in the domain of the unbounded
operator $\hat{A}$, which is not always the case. In fact, the Robertson uncertainty relation is false if $\hat{A}$ is an angle variable and $\hat{B}$ is the derivative with respect to this variable. In this example, the commutator is a nonzero constant-just as in the Heisenberg uncertainty relation and yet there are states where the product of the uncertainties is zero.  This issue can be overcome by using a variational method for the proof,  or by working with an exponentiated version of the canonical commutation relations.

Note that in the general form of the Robertson-Schr\"{o}dinger uncertainty relation, there is no need to assume that the operators $\hat{A}$ and $\hat{B}$ are self-adjoint operators. It suffices to assume that they are merely symmetric operators.  Since the Robertson and Schr\"{o}dinger relations are for general operators, the relations can be applied to any two observables to obtain specific uncertainty relations.

\subsection{The error-disturbance uncertainty relation}
The error-disturbance uncertainty relation (EDR) is one of the most fundamental issues in quantum mechanics since the EDR describes a peculiar limitation on measurements of quantum mechanical observables.

Ozawa's inequality \cite{27,28,29,30} as a correction is
$$\epsilon(A) \eta(B)+\epsilon(A) \sigma(B)+\sigma(A) \eta(B) \geq \frac{|\langle\psi|[\hat{A},\hat{B}]_{QPB}| \psi\rangle|}{2}$$

Branciard \cite{31} has improved Ozawa's relation-error-disturbance relation as
\begin{align}
&\left(\epsilon(A)^{2} \sigma(B)^{2}+\sigma(A)^{2} \eta(B)^{2}\right.\\
&+2 \epsilon(A) \eta(B) \sqrt{\sigma(A)^{2} \sigma(B)^{2}-C^{2}})^{\frac{1}{2}} \geq C \notag
\end{align}
where ${{C}}=\left| \left\langle {{\left[ \hat{A},\hat{B} \right]}_{QPB}} \right\rangle  \right|/2$.  In conclusively,  there are four situations given by \cite{33}
\begin{description}
  \item[i] The generalized form of Heisenberg's EDR for an arbitrary pair of observables $\hat{A}$ and $\hat{B}$ is given by
    $\epsilon\left( A \right)\eta \left( B \right)\ge C$.

  \item[ii]  It should be
emphasized that case {\bf{i}} is not equivalent to the following
relation that is mathematically proven
$\sigma \left( A \right)\sigma \left( B \right)\ge C$,
where $\sigma \left( A \right)$ is the standard deviation.
Indeed, Heisenberg's EDR in case {\bf{i}} is derived from case {\bf{ii}} under
certain additional assumptions.

  \item[iii] Ozawa proposed an alternative EDR that
is theoretically proven to be universally valid:
 $$ \epsilon(A) \eta(B)+\epsilon(A) \sigma(B)+\sigma(A) \eta(B) \geq C$$
  The presence of two additional terms indicates that the
first Heisenberg's term $\epsilon\left( A \right)\eta \left( B \right)$ is allowed to be lower than $C$.
   \item[iiii]
  Most recently, Branciard has improved Ozawa's
EDR as
\begin{align}
&\left(\epsilon^{2}(A) \sigma^{2}(B)+\sigma^{2}(A) \eta^{2}(B)\right.
+2 \epsilon(A) \eta(B) \sqrt{\sigma^{2}(A) \sigma^{2}(B)-C^{2}})^{\frac{1}{2}} \geq C\notag
\end{align}
  which is universally valid and tighter than Ozawa's EDR.
\end{description}
Obviously, it's clear to see that four situations completely rely on the commutator of operators $\hat{A}$ and $\hat{B}$, which is their common ground, it's a key point for all inequalities shown above.  All inequalities are still the superior approximate estimations, and a precise estimation to the real world is needed in science.

Actually, up to now, all such studies about the error-disturbance uncertainty relation or inequalities related to uncertainty relation or the generalized uncertainty relation are obtained by using the QPB or the commutator.
 Throughout all the current theoretical forms related to the uncertainty relation or correlative inequalities, environmental variables have never been taken into account so far, this point can be easily seen from the above inequalities of different forms.

As a result, the QCPB should be a substitute for all the QPB as a core in the error-disturbance uncertainty relation,
\begin{align}
 C &=\left| \left\langle \left[ \hat{A},\hat{B} \right] \right\rangle  \right|/2=\left| \left\langle {{\left[ \hat{A},\hat{B} \right]}_{QPB}}+G\left( s,\hat{A},\hat{B} \right) \right\rangle  \right|/2 \notag
\end{align}
As it shown, the quantum geometric bracket $G\left( s,\hat{A},\hat{B} \right)$ occupies a large proportion in the formula which contains the environment variable.

\subsection{Main results}
This work is inspired by the idea of the QCPB in \cite{15}.
As a consequence of the discovery of the QCPB and its related theory framework, the classic QPB is in the affirmative to be replaced by the QCPB theory \[{{\left[\hat{A},\hat{B} \right]}_{QPB}}\to \left[ \hat{A},\hat{B}\right]={{\left[ \hat{A},\hat{B}\right]}_{QPB}}+G\left( s,\hat{A},\hat{B} \right)\]Such replacement clearly is a natural method to reconsider the Heisenberg uncertainty relation and its development, but how to be compatible with such replacement is the key point.  This raises the compelling question: what will happen to the Heisenberg uncertainty relation under the QCPB, owing to completeness of the QCPB,  a natural answer would be Heisenberg uncertainty relation that is not complete according to the view of Einstein, it needs to be revised to an equality form.
That the search for such an equality might be a fruitful route
to prove that quantum mechanics is not a complete theory.
According to the QCPB, such an equality form will be a new physical interpretation of quantum measurement and the detailed information now
available about it.

We ask if there is an independent equality for deeply explaining such inequality. The aim is to find such certainly equality for better realizing this quantum inequality.   The discovery of the QCPB theory implies that there surely exists a problem of the uncertainty relation associated with the QPB.  Our goal in this paper is to take structure function $s$ generated by the spacetime or the environment into considerations, and we give a result as follows.

{\bf{Quantum geomertainty relation (QGR):}}
The quantum operator $\hat{f},~\hat{g}$ satisfies the QCPB which does not covariant commute,  then the standard variance between them is
  \[\Delta f\cdot \Delta g=\varpi\left( s,\hat{f},\hat{g} \right)\]
  where $s$ is the environment variable and $\varpi\left( s,\hat{f},\hat{g} \right)$ is quantumetric function.

Our method for building the quantum geomertainty relation (QGR) is compatible to be based on the QCPB  \cite{15} that is a complete description of reality.  Note that the structure function $s$ in QCPB is directly produced by the environment as a geometric structure function, hence, structure function $s$ in QCPB can be regarded as a environment variable,  the precise expression of the structure function $s$ is determined by the space we consider, hence, now the environment is naturally included in a complete formula which can be a correction term for the uncertainty relation, and then to make uncertainty relation certain is inevitable in such new formula associated with the QCPB.
In \cite{15}, the QCPB can completely explain the evolution behavior of all observable measurements of a physical system.   This paper is affirmatively built on the strong foundation of \cite{15} to completely explain the evolution behavior of observables without any uncertainty or randomness.

With the indeterminism of observable quantities constrained by the uncertainty principle. The question arises whether there might be some deeper reality hidden beneath quantum mechanics, to be described by a more
fundamental theory that can always predict the outcome of each measurement with certainty, this paper by using the quantum covariant Poisson bracket attempts to answer this question. Meanwhile, the uncertainty relation is just a derivation from this quantum geometric certainty equality. It deals with the measurement in quantum equality for different manifolds equipped with various mathematical or physical structure. Accordingly, the environment joins the physical process,  by taking environment variable as a geometric structure function in the quantum covariant Poisson bracket into consideration, it has naturally solved the environment problem for the measurements.  Doubtlessly, the quantum covariant Poisson bracket is surely a new way for such complete description of reality.

\section{Generalized geometric commutator and geomutator}

Let $a$ and $b$ be any two operators, their commutator is formally defined by
${\displaystyle [a,b]_{cr}=ab-ba.}$
Note that the operator for commutator can be any mathematical form to be appeared for the calculation, such as a function, vector, differential operator, partial differential operator, even a number in a number field, and so on, it can be arbitrarily chosen according to our needs.

Let $a$ and $b$ be any elements of any algebra, or any operators, their generalized geometric commutator is formally defined by the following.
\begin{definition}[GGC]\label{d1}
A generalized geometric commutator (GGC) of arity two $a,b$ is formally given by
$$\left[ a,b \right]={{\left[ a,b \right]}_{cr}}+G\left( s,a,b \right)$$
The geomutator is $$G\left( s,a,b \right)=a{{\left[ s,b \right]}_{cr}}-b{{\left[ s,a \right]}_{cr}}$$satisfying $G\left( s,a,b \right)=-G\left( s,b,a\right)$, where $s$ is a geometric structure function or geometric potential function given by domain.
\end{definition}

Some properties of the geomutator are given by
\begin{align}
  & G\left( s,a+b,c+d \right)=G\left( s,a,c \right)+G\left( s,b,d \right)+G\left( s,b,c \right)+G\left( s,a,d \right) \notag\\
 & G\left( s,a+b,c \right)=G\left( s,a,c \right)+G\left( s,b,c \right)\notag \\
 & G\left( s,a,c+d \right)=G\left( s,a,c \right)+G\left( s,a,d \right)\notag
\end{align}
for operators $a,b,c,d$. With some particular properties are given by
\begin{align}
  & G\left( s,a,a \right)=G\left( s,s,s \right)=0 \notag\\
 & G\left( s,s,a \right)=s{{\left[ s,a \right]}_{cr}} \notag\\
 & G\left( s,a,s \right)=s{{\left[ a,s \right]}_{cr}}\notag
\end{align}

\subsection{QCPB and quantum geobracket (QGB)}
As \cite{15} stated, quantum covariant Poisson bracket (QCPB) is defined by generalized geometric commutator (GGC) while quantum geometric bracket (QGB) is given based on the geomutator.  More precisely,

\begin{definition}[QCPB]\cite{15}
  The QCPB is generally defined as
 \[\left[ \hat{f},\hat{g} \right]={{\left[ \hat{f},\hat{g} \right]}_{QPB}}+G\left( s,\hat{f},\hat{g} \right)\]in terms of quantum operator $\hat{f},~\hat{g}$, where $$G\left(s, \hat{f},\hat{g} \right)=\hat{f}{{\left[ s,\hat{g} \right]}_{QPB}}-\hat{g}{{\left[ s,\hat{f} \right]}_{QPB}}=-G\left(s, \hat{g},\hat{f} \right)$$ is called quantum geometric bracket,  where $s$ is geometric potential function or the structure function.
\end{definition}It is zero if and only if $\hat{f}$ and $\hat{g}$ covariant commute, i,e. $\left[ \hat{f},\hat{g} \right]=0$.  It is remarkable to see that the QCPB representation admits a dynamical geometric bracket formula on the manifold. Note that structural function $s$ represents the background property of spacetime.

It is believed that quantum mechanics is an incomplete description of reality.
The QCPB suggests that hidden variables of certain types exist for a complete description of physical reality. In particular, the quantum geometric bracket (QGB) which means that there exsits hidden variables in the complete description such that quantum mechanics is complete.

Based on the QCPB theory, the fact that we may accurately predict some of the observations in quantum mechanics that can be linked to the this known hidden variables in the QGB--environment variable. Once this hidden variable
is determined, any observable measurement can be given accurately. As a complete hidden variable theory, there is no doubt that its results can return to the results given by quantum mechanics under
certain conditions, and it can obviously predict some different results from quantum mechanics shown in \cite{15}.  Later, in this way, we will know that this environment variable in the QGB can meet the above previously requirements of determinism.

Let's assert the role of the structure function $s$ that is a geometric structure function given by domain based on the generalized geometric commutator in definition \ref{d1}, it means that structure function $s$  is only determined by the environment, or spacetime, or manifolds, the domain, ect, from this viewpoint, the environment joins the physical process, the influence of the environment now based on new theory can't be ignored, it's naturally necessary to be considered in a physical process.  In other words, the environment has joined the physical process, accordingly. By using the antisymmetric, the quantum geometric bracket (QGB) is rewritten as
\[G\left( s,\hat{f},\hat{g} \right)=\hat{f}{{\left[ s,\hat{g} \right]}_{QPB}}+\hat{g}{{\left[ \hat{f},s \right]}_{QPB}}\]
This expression can vividly and concretely state that the environment interacts with the operators $\hat{f},\hat{g}$ respectively. We can say that the environment variable $s$ can be such hidden variable.

\begin{definition}\cite{15}
 The covariant equilibrium equation is given by $\left[ \hat{f},\hat{g} \right]=0$, i.e,
  \[{{\left[ \hat{f},\hat{g} \right]}_{QPB}}+G\left( s, \hat{f},\hat{g} \right)=0\]for operators $\hat{f},~\hat{g}$.
\end{definition}
Taking the modulus of means of the both sides
\[\left| \left\langle G\left( s,\hat{f},\hat{g} \right) \right\rangle  \right|=\left| \left\langle {{\left[ \hat{f},\hat{g} \right]}_{QPB}} \right\rangle  \right|\]
where $G\left( s,\hat{f},\hat{g} \right)=\hat{f}{{\left[ s,\hat{g} \right]}_{QPB}}+\hat{g}{{\left[\hat{f},s \right]}_{QPB}}$,   and applying the triangular inequality, we have
\[\left| \left\langle {{\left[ \hat{f},\hat{g} \right]}_{QPB}} \right\rangle  \right|\leq \left| \left\langle \hat{f}{{\left[ s,\hat{g} \right]}_{QPB}} \right\rangle  \right|+\left| \left\langle \hat{g}{{\left[ \hat{f},s \right]}_{QPB}} \right\rangle  \right|\]

\subsection{Covariant dynamics and G-dynamics}
In this section, we will briefly review the entire theoretical framework of quantum covariant Hamiltonian system defined by the quantum covariant Poisson bracket totally based on the paper \cite{15}.  More precisely,
the time covariant evolution of any observable $\hat{f}$ in the covariant dynamics is given by $$\frac{\mathcal{D}\hat{f}}{dt}=\frac{1}{\sqrt{-1}\hbar }\left[ \hat{f},\hat{H} \right]$$ It contains two dynamics:

The generalized Heisenberg equation of motion:

$\frac{d\hat{f}}{dt}=\frac{1}{\sqrt{-1}\hbar }{{\left[ \hat{f},\hat{H} \right]}_{QPB}}-\frac{1}{\sqrt{-1}\hbar }\hat{H}{{\left[ s,\hat{f} \right]}_{QPB}}$.

The G-dynamics: $\hat{w}=\frac{1}{\sqrt{-1}\hbar }{{\left[ s,\hat{H} \right]}_{QPB}}$,\newline
where $\frac{\mathcal{D}}{dt}=\frac{d}{dt}+\hat{w}$ is covariant time operator formally, and
${{\hat{H}}}$ is the Hamiltonian and $[\cdot,\cdot]$ denotes the GGC of two operators.

With the help of the structural function $s$ or the environment variable, the QCPB is well-defined for covariant quantum mechanics\footnote{Notes: GCC: Geometric canonical commutation; CCHE: canonical covariant Hamilton equations; CD: covariant dynamics;   GHE: generalized Heisenberg equations;  CTHE: Canonical thorough Hamilton equations.}.
\begin{center}
  \begin{tabular}{c r @{.} l}
\hline\hline
covariant quantum mechanics  \\
\hline
QCPB: $\left[ \hat{f},\hat{g} \right]={{\left[ \hat{f},\hat{g} \right]}_{QPB}}+G\left(s,\hat{f},\hat{g} \right)$, $G\left(s,\hat{f},\hat{g} \right)=\hat{f}{{\left[ s,\hat{g} \right]}_{QPB}}-\hat{g}{{\left[ s,\hat{f} \right]}_{QPB}}$ \\
\hline
GCC: $\left[ x_{i},{{{\hat{p}}}_{j }}\right]=\sqrt{-1}\hbar{{D}_{j}}{{x}_{i}}$, ${{D}_{i}}={{\partial }_{i}}+{{\partial }_{i}}s,{{\partial }_{i}}=\frac{\partial }{\partial {{x}_{i}}}$. \\
\hline
 Hamiltonian : $\hat{H}$  \\
\hline
CD:  $\frac{\mathcal{D}}{dt}{{x}_{i }}=\frac{1}{\sqrt{-1}\hbar }\left[ {{{x}}_{i}},\hat{H} \right],~~~\frac{\mathcal{D}}{dt}{{\hat{p}}_{j }}=\frac{1}{\sqrt{-1}\hbar }\left[ {{{\hat{p}}}_{j }},\hat{H} \right]$
\\
\hline
GHE: ${\dot{{x}_{i}}}=\frac{1}{\sqrt{-1}\hbar }{{\left[ {{{x}}_{i }},\hat{H} \right]}_{QPB}},~~\overset{\cdot }{\mathop{{{{\hat{p}}}_{j}}}}\, =\frac{1}{\sqrt{-1}\hbar }\left( {{\left[ {{{\hat{p}}}_{j }},\hat{H} \right]}_{QPB}}-\hat{H}\left[ s,{{{\hat{p}}}_{j }} \right]_{QPB} \right)$
\\
\hline
The G-dynamics: $\hat{w}=\frac{1}{\sqrt{-1}\hbar }{{\left[ s,\hat{H} \right]}_{QPB}}$
\\
\hline\hline
\end{tabular}
\end{center}
where $\left[ \hat{f},\hat{g} \right]_{QPB}=\hat{f}\hat{g}-\hat{g}\hat{f}$ is quantum Poisson brackets for operators.

Covariant dynamics, generalized Heisenberg equation, G-dynamics are strong evidences to affirmatively say a fact that quantum mechanics is an incomplete theory. It strongly reminds of us that there exactly exists an undiscovered theory associated with the environment factor we still seek behind quantum mechanics, which can completely explain the evolution behavior of all observable measurements of the physical system and it can naturally avoid any uncertainty or randomness based on the covariant dynamics, in particular, the existence of G-dynamics unquestionably supports this perspective.

\subsection{Geomentum operator}
\begin{definition}
  Let $M$ be a smooth manifold represented by geometric potential function $s$, then geomentum operator is
  $\hat{p}=-\sqrt{-1}\hbar D$, where $D=\nabla +\nabla s$. The component is ${{\hat{p}}_{j}}=-\sqrt{-1}\hbar {{D}_{j}}$ in which ${{D}_{j}}={{\partial }_{j}}+{{\partial }_{j}}s$ holds.
\end{definition}
Note that the geomentum operator is a revision of the classical momentum operator.

\begin{theorem}[Geometric canonical quantization rules]
  Geometric equal-time canonical commutation relation is
  \[\left[ {{x}_{i}},\hat{{{p}_{j}}} \right]=\sqrt{-1}\hbar {{D}_{j}}{{x}_{i}}\]where  $\left[ \cdot ,~\cdot  \right]={{\left[ \cdot ,~\cdot  \right]}_{QPB}}+G\left(s,\cdot ,~\cdot  \right)$ is QCPB.
\end{theorem}
Geometric canonical commutation relation can be expressed in a specific form \[\left[ {{{x}_{i}}},\hat{{{p}_{j}}} \right]=\sqrt{-1}\hbar \left( {{\delta }_{ij}}+{{x}_{i}}\frac{\partial }{\partial {{x}_{j}}}s \right) \]
In other words, it also can be rewritten as
$\left[ {{x}_{i}},\hat{{{p}_{j}}} \right]=\sqrt{-1}\hbar {{\theta }_{ij}}$, where
${{\theta}_{ij}}={{\delta }_{ij}}+{{x}_{i}}{{\partial }_{j}}s$.   On the basis of the geometric equal-time canonical commutation relation above, we can further evaluate it.
\begin{align}
{{\partial }_{k}}\left[ {{x}_{i}},\hat{{{p}_{j}}} \right]  &=\sqrt{-1}\hbar \frac{\partial }{\partial {{x}_{k}}}\left( {{\delta }_{ij}}+{{x}_{i}}\frac{\partial s}{\partial {{x}_{j}}} \right) \notag\\
 & =\sqrt{-1}\hbar {{x}_{i}}\frac{{{\partial }^{2}}s}{\partial {{x}_{k}}\partial {{x}_{j}}}+\sqrt{-1}\hbar {{\delta }_{ik}}\frac{\partial s}{\partial {{x}_{j}}} \notag
\end{align}
where ${{\delta }_{ik}}=\frac{\partial }{\partial {{x}_{k}}}{{x}_{i}}$. Accordingly, it yields
\begin{align}\label{eq4}
 {{D}_{k}}\left[ {{x}_{i}},\hat{{{p}_{j}}} \right] &=\sqrt{-1}\hbar \left( {{x}_{i}}{{Q}_{kj}}+{{\delta }_{ik}}\frac{\partial s}{\partial {{x}_{j}}}+{{\delta }_{ij}}{{A}_{k}} \right)
\end{align}
where ${{Q}_{kj}}=\frac{\partial {{A}_{j}}}{\partial {{x}_{k}}}+A_{k}A_{j}$ can be seen as the second order curvature tensor, where $A_{j}=\frac{\partial s}{\partial {{x}_{j}}}$.
As we see, this further derivation shows that the environment variable has
abundant implication given to be inevitable for the quantum mechanics, \eqref{eq4}
is only associated with the structure function $s$.
For this reason, it can be regarded as a hidden variable undiscovered all the time until the discovery of the QCPB.

\subsection{Geometric anticommutator (GAC) and anti-geomutator}
Obviously, in some sense, the anticommutator also needs to be generalized as the commutator does,  Analogously, denote by ${{\left\{ a,b \right\}}_{ir}}=ab+ba$ the anticommutator,
\begin{definition}
 The geometric anticommutator (GAC) of any two elements $a$ and $b$ is defined by
  \[\left\{ a,b \right\}={{\left\{ a,b \right\}}_{ir}}+Z\left(s, a,b \right)\]
  where $Z\left(s,a,b \right)=Z\left(s, b,a\right)=a{{\left\{ s,b \right\}}_{ir}}+b{{\left\{ s,a \right\}}_{ir}} $ is called anti-geomutator, $s$ is geometric potential function.
\end{definition}
As the definition above stated,
the anti-geomutator can be taken as
\[Z\left(s,a,b \right)=\left( a:s:b \right)+{{\left\{ a,b \right\}}_{ir}}s\]
where $\left( a:s:b \right)=asb+bsa$, and $s$ is the geometric function created by the environment. As seen, this form is very similar to the quantum geometric bracket, this is why we need to generalize the anticommutator.

Note that in order to preserve the related property of the anticommutator, we always let anti-geomutator $Z\left(s, a,b \right)$ satisfiy the symmetry, namely,  $Z\left(s,a,b \right)=Z\left(s, b,a\right)$.
As a result of the symmetry of anti-geomutator, geometric anticommutator then follows the symmetry $\left\{ a,b \right\}=\left\{ b,a \right\}$.

In conclusions, the generalized geometric commutator (GGC) and the geometric anticommutator (GAC) can be listed as below
\begin{align}
  & \left[a,b \right]={{\left[ a,b \right]}_{cr}}+G\left( s,a,b \right), \notag\\
 & \left\{ a,b \right\}={{\left\{ a,b \right\}}_{ir}}+Z\left( s,a,b \right), \notag
\end{align}
where $G\left( s,a,b \right)=a{{\left[ s,b \right]}_{cr}}-b{{\left[ s,a \right]}_{cr}}$ and $Z\left( s,a,b \right)=a{{\left\{ s,b \right\}}_{ir}}+b{{\left\{ s,a \right\}}_{ir}}$.

When the $a,b$ come to the operators $\hat{A},\hat{B}$ respectively, then
\begin{align}
  & \left\{ \hat{A},\hat{B} \right\}={{\left\{ \hat{A},\hat{B} \right\}}_{ir}}+\left( \hat{A}:s:\hat{B} \right)+{{\left\{ \hat{A},\hat{B} \right\}}_{ir}}s \notag\\
 & \left[ \hat{A},\hat{B} \right]={{\left[ \hat{A},\hat{B} \right]}_{QPB}}+\left\langle \hat{A}:s:\hat{B} \right\rangle -{{\left[ \hat{A},\hat{B} \right]}_{QPB}}s \notag
\end{align}
If the operators do not commute:
$\left[{\hat  {A}},{\hat  {B}}\right]_{QPB}\psi \neq 0$,
they can't be prepared simultaneously to arbitrary precision, and there is an uncertainty relation between the observables,
$ {\displaystyle \Delta A\Delta B\geq \left|{\frac {\langle [{\hat {A}},{\hat {B}}]_{QPB}\rangle }{2}}\right|}$,
even if $\psi$ is an eigenfunction the above relation holds. Notable pairs are position and momentum, and energy and time-uncertainty relations, and the angular momenta (spin, orbital and total) about any two orthogonal axes.

All such nontrivial commutation relations for pairs of operators lead to corresponding uncertainty relations, involving positive semi-definite expectation contributions by their respective commutators and anticommutators. In general, for two Hermitian operators ${\hat {A}}$ and ${\hat {B}}$, consider expectation values in a system in the state $\psi$, the variances around the corresponding expectation values being ${{\left( \Delta A \right)}^{2}}=\left\langle {{\left( {\hat {A}}-\left\langle {\hat {A}} \right\rangle  \right)}^{2}} \right\rangle $, etc.

Then
$$ \Delta  A \, \Delta  B \geq  \frac{1}{2} \sqrt{ \left|\left\langle\left[{{\hat {A}}},{{\hat {B}}}\right]_{QPB}\right\rangle \right|^2 + \left|\left\langle\left\{ {\hat {A}}-\langle {\hat {A}}\rangle ,{\hat {B}}-\langle {\hat {B}}\rangle  \right\}_{ir} \right\rangle \right|^2}$$
 This follows through use of the Cauchy-Schwarz inequality, since $$\left| \left\langle {{{\hat {A}}}^{2}} \right\rangle \left\langle {{{\hat {B}}}^{2}} \right\rangle  \right|\ge {{\left| \left\langle {\hat {A}}{\hat {B}} \right\rangle  \right|}^{2}}$$ and ${\hat {A}}{\hat {B}}=\left( {{\left[ {\hat {A}},{\hat {B}} \right]}_{QPB}}+{{\left\{ {\hat {A}},{\hat {B}}\right\}}_{ir}} \right)/2$,  and similarly for the shifted operators ${\hat {A}}-\left\langle {\hat {A}}\right\rangle $ and ${\hat {B}}-\left\langle {\hat {B}}\right\rangle$.
Substituting for ${\hat {A}}$ and ${\hat {B}}$ yield Heisenberg's familiar uncertainty relation for $x$ and $\hat {p}$, as usual.

As quantum mechanics proven, there is a theorem for such uncertainty relation given by
\begin{theorem}\label{t2}
If $\hat{A},\hat{B},\hat{C},\hat{D}$ are real Hermitian operators satisfying the following relation
$\sqrt{-1}\hat{C}={{\left[ \hat{A},\hat{B} \right]}_{QPB}}$ and
$\hat{D}={{\left\{ \hat{A},\hat{B} \right\}}_{ir}}$,
then it has
\[\overline{{{\hat{A}}^{2}}}\cdot \overline{{{\hat{B}}^{2}}}\ge \frac{1}{4}\left[ {{\left( \overline{\hat{C}} \right)}^{2}}+{{\left( \overline{\hat{D}} \right)}^{2}} \right]\]
\end{theorem}
Obviously, according to the GGC and the GAC, the theorem \ref{t2} accordingly needs to be revised to fit the quantum reality.  On the foundation of these new theories, we attempt to search a complete theory for the quantum mechanics, thusly, to begin with the following logical derivation.

In this paper, we always assume that if for any Hermitian operator ${\hat {A}}$ that is not an angle variable and ${\hat {B}}$ is not the derivative with respect to this variable.
And the vector $\hat{B}\left| \psi  \right\rangle $ is in the domain of the unbounded operator $\hat{A}$ in the most situations. Then based upon the definition of variance, we have a definition for linear transformation given by
\begin{definition}\cite{4}\label{d11}
  Let $\hat{X}$ be any operator, then geoperator can be defined as
  \[{{\hat{X}}^{\left( s \right)}}=\hat{X}+\lambda\left( s,\hat{X}\right)\]
  where $\lambda\left( s,\hat{X}\right)={{e}^{-s}}\hat{X}{{e}^{s}}-\hat{X}$ is coupling interaction between the observable $\hat{X}$ and the environment $s$. If
  ${{\left[ s,\hat{X} \right]}_{QPB}}=0$, then $\lambda\left( s,\hat{X}\right)=0$.
\end{definition}
We denote $$\lambda\left( s,\hat{X}\right)=-{{\left[ s,\hat{X} \right]}_{QPB}}+\frac{1}{2!}{{\left[ s,{{\left[ s,\hat{X} \right]}_{QPB}} \right]}_{QPB}}-\frac{1}{3!}{{\left[ s,{{\left[ s,{{\left[ s,\hat{X} \right]}_{QPB}} \right]}_{QPB}} \right]}_{QPB}}+\cdots$$It's obvious to see that non-trivial result $$\lambda\left( s,\hat{X}\right)={{e}^{-s}}\hat{X}{{e}^{s}}-\hat{X}$$ holds for all operator $\hat{X}$.  Therefore, some special results follow $$\lambda\left( Constants,\hat{X}\right)=\lambda\left( 0,\hat{X}\right)=\lambda\left( s,x\right)=0$$ where $x$ is the position operator. General speaking, $\lambda\left( s,\hat{X}\right)\neq 0$ holds for the non-trivial cases. When this situation goes to the classical momentum operator
\[{{\hat{p}}^{\left( ri \right)}}={{e}^{-s}}{{\hat{p}}^{\left( cl \right)}}{{e}^{s}}={{\hat{p}}^{\left( cl \right)}}-{{\left[ s,{{\hat{p}}^{\left( cl \right)}} \right]}_{QPB}}={{\hat{p}}^{\left( cl \right)}}+{{{\hat{p}}}^{\left( cl \right)}}s\]It generates the geomentum operator,
where $$\sqrt{-1}\hbar u={{\left[ s,{{{\hat{p}}}^{\left( cl \right)}}\left( t \right) \right]}_{QPB}}=-{{{\hat{p}}}^{\left( cl \right)}}s\neq0$$ and ${{\left[ s,{{\left[ s,{{\hat{p}}^{\left( cl \right)}} \right]}_{QPB}} \right]}_{QPB}}=0$ holds, the high order terms disappear

Subsequently, the transformation of the mean error of mechanical quantity operator $\hat{X}$ follows
\[\Delta \hat{X}\to \Delta {{\hat{X}}^{\left( s \right)}}={{\hat{X}}^{\left( s \right)}}-\left\langle {{\hat{X}}^{\left( s \right)}} \right\rangle =\Delta \hat{X}+\Delta I\]where $\Delta I=\lambda\left( s,\hat{X}\right)-\left\langle \lambda\left( s,\hat{X}\right) \right\rangle$. For examples, if $\hat{X}=x$ is a position operator, then $\Delta I=0$ which means $\Delta {{\hat{X}}^{\left( s \right)}}=\Delta \hat{X}$. If $\hat{X}={{{\hat{p}}}^{\left( cl \right)}}$ is the classical momentum operator, then $\Delta I={{{\hat{p}}}^{\left( cl \right)}}s-\left\langle {{{\hat{p}}}^{\left( cl \right)}}s\right\rangle$

Obviously, the geometric operator is a extension of the Hermitian operator. the geometric operator can be a non-Hermitian operator,
Note that the interaction term $\lambda\left( s,\hat{X}\right)\neq0$ actually is a precise operator with respect to the spacetime, as we state, the structure function $s$ generated by the space or manifolds as an environment variable only associate with the space or the manifolds, and it's independent to the wave function.

Note that the definition of the geoperator is reasonable, in particular, the evidence comes from the coupling interaction between the observable $\hat{X}$ and the environment variable $s$ that is described by $\lambda\left( s,\hat{X}\right)$ in which the environment variable $s$. In fact, geomentum operator is one of the geoperator if the Hermitian operator $\hat{X}=\hat{p}^{\left( cl \right)} = -\sqrt{-1}\hbar\nabla$.

\subsection{The QPB and the QCPB}

In this section, we reconsider the uncertainty relation and give an full equality interpretation of such uncertainty relation based on the QCPB.
As stated, the QCPB is a covariant form which links to the environment represented by the structure function. For this strong reason, it means that uncertainty relation can be further developed to be a certainty identity.
The QCPB implies that it is in general possible to predict the value of a quantity with arbitrary certainty, even if all initial conditions are specified.

To start with a difference calculation between the QCPB and the QPB,
\begin{align}
 \left| \frac{\overline{\left[ \hat{f},\hat{g} \right]}}{2\sqrt{-1}} \right|-\left| \frac{\overline{{{\left[ \hat{f},\hat{g} \right]}_{QPB}}}}{2\sqrt{-1}} \right| & =\frac{1}{2}\left( \left| \overline{\left[ \hat{f},\hat{g} \right]} \right|-\left| \overline{{{\left[ \hat{f},\hat{g} \right]}_{QPB}}} \right| \right) \notag\\
 & =\frac{1}{2}\left( \left| \overline{{{\left[ \hat{f},\hat{g} \right]}_{QPB}}}+\overline{\hat{f}{{\left[ s,\hat{g} \right]}_{QPB}}}-\overline{\hat{g}{{\left[ s,\hat{f} \right]}_{QPB}}} \right|-\left| \overline{{{\left[ \hat{f},\hat{g} \right]}_{QPB}}} \right| \right) \notag
\end{align}
Applying the inequality to above computation, and it yields \[\left| \overline{{{\left[ \hat{f},\hat{g} \right]}_{QPB}}}+\overline{G\left( s,\hat{f},\hat{g} \right)} \right|-\left| \overline{{{\left[ \hat{f},\hat{g} \right]}_{QPB}}} \right|\le \left| \overline{G\left( s,\hat{f},\hat{g} \right)} \right|\]
Obviously, we can see that there appears upper bound of differnce which implies an existence of certainty identity as a correction of the uncertainty relation.
As a result,  we obtain
\[\left| \frac{\overline{\left[ \hat{f},\hat{g} \right]}}{2\sqrt{-1}} \right|-\left| \frac{\overline{{{\left[ \hat{f},\hat{g} \right]}_{QPB}}}}{2\sqrt{-1}} \right|\le \frac{1}{2}\left| \overline{G\left( s,\hat{f},\hat{g} \right)} \right|\]
Actually, it states the inequality of the expectation value of the QCPB
\[\left| \overline{\left[ \hat{f},\hat{g} \right]} \right|=\left| \overline{{{\left[ \hat{f},\hat{g} \right]}_{QPB}}+G\left( s,\hat{f},\hat{g} \right)} \right|\le \left| \overline{{{\left[ \hat{f},\hat{g} \right]}_{QPB}}} \right|+\left| \overline{G\left( s,\hat{f},\hat{g} \right)} \right|\]Meanwhile, absolute value of the expectation value of the quantum geometric bracket also has a restriction expressed as
\[ \left| \overline{G\left( s,\hat{f},\hat{g} \right)} \right|\le \left| \overline{\hat{f}{{\left[ s,\hat{g} \right]}_{QPB}}} \right|+\left| \overline{\hat{g}{{\left[ s,\hat{f} \right]}_{QPB}}} \right|\]
Accordingly, \[\left| \overline{\left[ \hat{f},\hat{g} \right]} \right|\le \left| \overline{\hat{g}{{\left[ s,\hat{f} \right]}_{QPB}}} \right|+\left| \overline{\hat{f}{{\left[ s,\hat{g} \right]}_{QPB}}} \right|+\left| \overline{{{\left[ \hat{f},\hat{g} \right]}_{QPB}}} \right|\]

As \eqref{eq1} shown, it turns out that uncertainty relation need to be revised,  in other words, it strongly implies that uncertainty relation is an approximate incomplete quantum theory. One thing we can ascertain is that all the QPB should be surely replaced by the complete theory--the QCPB theory. It means that $\left\langle  \psi  \right|{{\left[\hat{A},\hat{B} \right]}_{QPB}}\left| \psi  \right\rangle \to \left\langle  \psi  \right|\left[\hat{A},\hat{B} \right]\left| \psi  \right\rangle$, in a precise expression,
\[\left\langle  \psi  \right|\left[\hat{A},\hat{B} \right]\left| \psi  \right\rangle =\left\langle  \psi  \right|{{\left[ \hat{A},\hat{B} \right]}_{QPB}}\left| \psi  \right\rangle +\left\langle  \psi  \right|G\left( s,\hat{A},\hat{B}\right)\left| \psi  \right\rangle \] for certain.
Clearly, the inequality is just a representation of the quantum geometric certainty identity.
More precisely, the QCPB in terms of operators $\hat{f}$ and $\hat{g}$ can show more details
\begin{align}
\left| \frac{\overline{\left[ \hat{f},\hat{g} \right]}}{2\sqrt{-1}} \right|&=\left| \overline{{{\left[ \hat{f},\hat{g} \right]}_{QPB}}}+\overline{\hat{f}{{\left[ s,\hat{g} \right]}_{QPB}}}-\overline{\hat{g}{{\left[ s,\hat{f} \right]}_{QPB}}} \right|/2 \notag
\end{align}
where $s$ is the environment variable.  Absolutely, the quantum geometric bracket  $G\left( s,\hat{f},\hat{g} \right)$  can be interpreted as the intrinsic interaction with the environment, it is precisely because of this that uncertainty occurs.  It needs to stress that the structure function $s$ represents the environment, it seems to the environment joins the measurement that is inevitable, all quantum process occur to this background.

\section{Quantum geomertainty relation}
In this section, we try to construct a new theory containing geometric potential function of physical reality for the reinterpretation of quantum mechanics.

Note that the QCPB does not covariant commute in terms of the operators $\hat{f},\hat{g}$ which means $\left[ \hat{f},\hat{g} \right]\ne 0$, more precisely, it says that inequality  ${{\left[ \hat{f},\hat{g} \right]}_{QPB}}+G\left( s,\hat{f},\hat{g} \right)\ne 0$ holds.

We propose such a geometric condition can be consistently obtained as a complete quantum theory of a solvable proposal.  Without loss of generality, we abstractly give the most general form of quantum geomertainty relation (QGR)

{\bf{ Quantum Geomertainty Relation (QGR):}}
The standard deviation for the operator  $\hat{A},\hat{B}$ satisfies the QCPB that does not covariant commute and then the quantum geomertainty relation (QGR) is
${{\sigma }_{A}}{{\sigma }_{B}}=\varpi\left( s,\hat{A},\hat{B} \right)$,
where  $s$ here can be treated as the environment variable, $\varpi=\varpi\left( s,\hat{A},\hat{B} \right)$ is quantumetric function.   The variance form is $\sigma _{A}^{2}\sigma _{B}^{2}=\wp \left( s,\hat{A},\hat{B} \right)$, then $\wp \left( s,\hat{A},\hat{B} \right)={{\varpi }^{2}}\left( s,\hat{A},\hat{B} \right)$.

Note that this abstract form is for a real reality, $\varpi\left( s,\hat{A},\hat{B} \right)$ and $\wp \left( s,\hat{A},\hat{B} \right)$ are real functions, obviously, as we see, the environment variable $s$ has joined the quantum process, and then it forms a complete system. The precise formula of the $\varpi\left( s,\hat{A},\hat{B} \right)$ or $\wp \left( s,\hat{A},\hat{B} \right)$ needs to seek to fit the physical truth, especially, to be compatible with the physical facts and experiments data are the factors to check a theory that is right or not.
Meanwhile, we're positively convinced that quantumetric function is a certain formula which contains the GGC and the GAC. Surely, the quantumetric function induce a related inequality in generally, in other words, there exists an inequality based on the quantumetric function for general case.  Based on the derivation of the geometric Schr\"{o}dinger uncertainty relation, we have adequate reasons to say that the inequality is just an approximate representation for the measurements of observables, contrarily,  it is convinced that the quantumetric function truly holds for a complete description for a new quantum mechanics and derives relevant inequality.

Note that for the quantum geomertainty relation (QGR), inspired by the QCPB and the GAC, we may decompose it to the two terms,
\[\varpi \left( s,\hat{A},\hat{B}  \right)={{\varpi }^{p}}\left( \hat{A},\hat{B}  \right)+{{\varpi }^{e}}\left( s,\hat{A},\hat{B} \right)\]
The first term ${{\varpi }^{p}}\left(\hat{A},\hat{B} \right)$ means the pure result from the both operators, the latter ${{\varpi }^{e}}\left( s,\hat{A},\hat{B}  \right)$ means the entangled term.

Note that the QGR indicates a fact always exists that the environment still interacts with the observables, the uncertainty actually comes from the ignorance of the environment, conversely, it says that a complete theory of quantum mechanics about the aspect of the measurement links to the environment, the abstract equality expressed by the QGR can well explain this problem. As a result of the QGR, we can see that the equality is expressed by two terms, all previous theories about the uncertainty relation only relates to the first term ${{\varpi }^{p}}\left( \hat{A},\hat{B} \right)$,  but now, we discover the complete commutator and anti-commutator, this further development for the  commutator and anti-commutator has naturally contained the environment, and then it potentially forms a complete theory to explain more secrets about the quantum world.

In favour of the quantum geomertainty relation (QGR) is the rigorous result for the quantum mechanics.  The concrete expression for the QGR is unknown, but we can refer to the derivation previously.  We will consider one case of QGR related to the revision of Schr\"{o}dinger uncertainty relation. To relate the uncertainty of the quantum mechanics with the environment of a variable in a space. This is a very good interpretation for understanding how quantum realm works actually.

\subsection{The generalized variance and geometric operator}

Let's always suppose that the Hermitian operator ${\hat {A}}$ is not an angle variable and ${\hat {B}}$ is not the derivative in terms of this angle variable in this paper.
And the $\hat{B}\psi $ is in the domain of the unbounded operator $\hat{A}$ in the most situations.

 Now, when we add the environment into the quantum system, then it's a complete system, the standard deviation between the operators in a state becomes certainty and complete equality. As a result, the quantum geomertainty relation (QGR) accordingly emerges.

Actually, the generalized variance $\Sigma _{X}^{2}$ in terms of the Hermitian operator $\hat{X}$ is a complete formula
\begin{align}
 \Sigma _{X}^{2} & =\overline{{{X}^{2}}}+\overline{{{\lambda}^{2}}}+\overline{\hat{X}\lambda+\lambda\hat{X}}-{{\overline{X}}^{2}}-{{\overline{\lambda}}^{2}}-2\bar{\lambda}\bar{X} =\overline{{{X}^{\left( s \right)}}^{2}}-{{\overline{{{X}^{\left( s \right)}}}}^{2}} \notag
\end{align}
for the reality in quantum mechanics.

\subsection{The QGR and entanglement term}
The QGR is given as follows.
\begin{theorem}\label{t5}
 The variance between the operators $\hat{A},\hat{B} $ in a state satisfies
 \[\sigma _{A}^{2}\sigma _{B}^{2}+\Theta \left( s,\hat{A},\hat{B} \right)=\Xi \left( s,\hat{A},\hat{B} \right)+\epsilon \left( s,\hat{A},\hat{B} \right)\]
where $\epsilon\left( s,\hat{A},\hat{B} \right)\geq0$, and geometric entanglement term is \[\Theta \left( s,\hat{A},\hat{B} \right)=\rho \left( A,s \right)\sigma _{B}^{2}+\sigma _{A}^{2}\rho \left( B,s \right)+\rho \left( A,s \right)\rho \left( B,s \right)\]and
\begin{align}
 & \Xi \left( s,\hat{A},\hat{B}\right)={{\left| \left( \frac{1}{2}\overline{\left\{  \hat{A},\hat{B} \right\}}+\theta \left( s,\hat{A},\hat{B}\right) \right) \right|}^{2}}+{{\left| \left( \frac{\overline{\left[  \hat{A},\hat{B} \right]}}{2\sqrt{-1}}+\vartheta \left( s,\hat{A},\hat{B} \right) \right) \right|}^{2}} \notag
\end{align}where $\left[ \hat{A},\hat{B} \right]$ and $\left\{ \hat{A},\hat{B} \right\}$ are respectively the GGC and the GAC in terms of operators $\hat{A},\hat{B}$, and

$\theta \left( s,\hat{A},\hat{B}\right)=\overline{uv}-\left( \left( \overline{A}+\overline{u} \right)\left( \overline{B}+\overline{v} \right) \right) $~~ $u=\lambda\left( \hat{A},s\right),~v=\lambda\left( \hat{B},s\right)$,

$\vartheta \left( s,\hat{A},\hat{B} \right)=-\sqrt{-1}\overline{{{\left[ \hat{A},\hat{B} \right]}_{QPB}}s}$, \\
and the entanglement term is $\rho \left( X,s \right)=\sigma _{u}^{2}-2\overline{u}\overline{X}+\overline{{{\left\{\hat{X},u \right\}}_{ir}}}$ for $\hat{X}=\hat{A},\hat{B}$.

\begin{proof}
 Based on the above definition, we have
  \[\Delta {{\hat{A}}^{\left( s \right)}}={{\hat{A}}^{\left( s \right)}}-\overline{{{\hat{A}}^{\left( s \right)}}}=\hat{A}-\overline{A}+u-\overline{u}=\Delta \hat{A}+\Delta u\]
where $\overline{\hat{A}}=\overline{A}$ has been used.  Then the deviation in terms of ${{\hat{A}}^{\left( s \right)}}$ writes
\begin{align}
{{\left( \Delta {{A}^{\left( s \right)}} \right)}^{2}}  & =\overline{{{\left( \Delta {{\hat{A}}^{\left( s \right)}} \right)}^{2}}}=\left( \psi ,{{\left( \Delta {{\hat{A}}^{\left( s \right)}} \right)}^{2}}\psi  \right)=\left( \psi ,{{\left( \hat{A}-\overline{A}+u-\overline{u} \right)}^{2}}\psi  \right) \notag\\
 & =\left( \psi ,{{\left( \Delta \hat{A}+\Delta u \right)}^{2}}\psi  \right)=\left( \psi ,\left( {{\left( \Delta \hat{A} \right)}^{2}}+\left( \Delta \hat{A} \right)\Delta u+\Delta u\left( \Delta \hat{A} \right)+{{\left( \Delta u \right)}^{2}} \right)\psi  \right) \notag\\
 & =\left( \psi ,{{\left( \Delta \hat{A} \right)}^{2}}\psi  \right)+\left( \psi ,\left( \left( \Delta \hat{A} \right)\Delta u \right)\psi  \right)+\left( \psi ,\Delta u\left( \Delta \hat{A} \right)\psi  \right)+\left( \psi ,{{\left( \Delta u \right)}^{2}}\psi  \right) \notag\\
 & =\overline{{{A}^{2}}}-{{\overline{A}}^{2}}+\left( \psi ,\left( \left( \Delta \hat{A} \right)\Delta u \right)\psi  \right)+\left( \psi ,\Delta u\left( \Delta \hat{A} \right)\psi  \right)+\left( \psi ,{{\left( \Delta u \right)}^{2}}\psi  \right)\notag \\
 & =\overline{{{A}^{2}}}-{{\overline{A}}^{2}}+\left( \psi ,\left( \hat{A}u-\overline{A}u-\hat{A}\overline{u}+2\overline{u}\overline{A}+u\hat{A}-\overline{u}\widehat{A}-u\overline{A}+{{u}^{2}}+{{\overline{u}}^{2}}-2u\overline{u} \right)\psi  \right) \notag\\
 & =\overline{{{A}^{2}}}-{{\overline{A}}^{2}}+\overline{Au}-2\overline{A}\overline{u}+2\overline{u}\overline{A}
 +\overline{uA}-2\overline{u}\overline{A}+\overline{{{u}^{2}}}-{{\overline{u}}^{2}} \notag\\
 & =\overline{{{A}^{2}}}-{{\overline{A}}^{2}}+\overline{Au}
 +\overline{uA}-2\overline{u}\overline{A}+\sigma _{u}^{2}\notag
\end{align}
where $\sigma _{u}^{2}=\overline{{{u}^{2}}}-{{\overline{u}}^{2}}$, and
\begin{align}
 {{\left( \hat{A}-\overline{A}+u-\overline{u} \right)}^{2}} &= {{\left( \Delta \hat{A} \right)}^{2}}+\left( \Delta \hat{A} \right)\Delta u+\Delta u\left( \Delta \hat{A} \right)+{{\left( \Delta u \right)}^{2}} \notag\\
 & ={{\hat{A}}^{2}}+{{\overline{A}}^{2}}-2\hat{A}\overline{A}+\hat{A}u-\overline{A}u \notag\\
 & -\hat{A}\overline{u}+2\overline{u}\overline{A}
 +u\hat{A}-\overline{u}\hat{A}-u\overline{A}
 +{{u}^{2}}+{{\overline{u}}^{2}}-2u\overline{u}\notag
\end{align}
More specifically, then the deviation in terms of ${A}^{\left( s \right)}$ rewrites
\[{{\left( \Delta {{A}^{\left( s \right)}} \right)}^{2}}=\sigma _{A}^{2}+\rho \left( A,s \right)=\sigma _{A}^{2}+\sigma _{u}^{2}+\overline{Au}+\overline{uA}-2\overline{u}\overline{A}\]
In order to distinguish the classical theories,
let \[\rho \left( A,s \right)=\overline{Au}+\overline{uA}-2\overline{u}\overline{A}
 +\sigma _{u}^{2}\]be denoted, then
${{\left( \Delta {{A}^{\left( s \right)}} \right)}^{2}}=\sigma _{A}^{2}+\rho \left( A,s \right)$.
Compactly, it's rewritten as \[{{\left( \Delta {{A}^{\left( s \right)}} \right)}^{2}}=\sigma _{A}^{2}+\rho \left( A,s \right)=\sigma _{A}^{2}+\sigma _{u}^{2}-2\overline{u}\overline{A}+\overline{{{\left\{ A,u \right\}}_{ir}}}\]
where \[\rho \left( A,s \right)=\sigma _{u}^{2}-2\overline{u}\overline{A}+\overline{{{\left\{ A,u \right\}}_{ir}}}\]can be interpreted as the interactions between the operator $\hat{A}$ and environment $s$.
Similarly, it has the expression for the operator $\hat{B}$ given by  \[{{\left( \Delta {{B}^{\left( s \right)}} \right)}^{2}}=\overline{{{B}^{2}}}-{{\overline{B}}^{2}}+\rho \left( B,s \right)\]
It should be noted that $v=\lambda\left( \hat{B},s\right)$ is for $\hat{B}$. In totally, \[\rho \left( X,s \right)=\overline{Xu}+\overline{uX}-2\overline{u}\overline{X}+\sigma _{u}^{2}\]
where $u=\lambda\left( \hat{X},s\right)$, and $\sigma _{u}^{2}=\overline{{{u}^{2}}}-{{\overline{u}}^{2}}$.
Furthermore,  the generalized variance in terms of ${{\hat{X}}^{\left( s \right)}}$ writes
\[\Sigma _{X}^{2}={{\left( \Delta {{X}^{\left( s \right)}} \right)}^{2}}=\sigma _{X}^{2}+\rho \left( X,s \right)=\sigma _{X}^{2}+\rho \left( X,s \right)\]
where entanglement term is \[\rho \left( X,s \right)=\sigma _{u}^{2}-2\overline{u}\overline{X}+\overline{{{\left\{\hat{X},u \right\}}_{ir}}}\]
where $u=\lambda\left( \hat{X},s\right)$, and $\hat{X}=\hat{A}, \hat{B}$.
 Thusly, we get
\begin{align}
  & {{\left( \Delta {{A}^{\left( s \right)}} \right)}^{2}}=\overline{{{A}^{2}}}-{{\overline{A}}^{2}}+\rho \left( A,s \right)={{\left( \Delta A \right)}^{2}}+\rho \left( A,s \right) \notag\\
 & {{\left( \Delta {{B}^{\left( s \right)}} \right)}^{2}}=\overline{{{B}^{2}}}-{{\overline{B}}^{2}}+\rho \left( B,s \right)={{\left( \Delta B \right)}^{2}}+\rho \left( B,s \right) \notag
\end{align}
Let ${{\left( \Delta A \right)}^{2}}=\sigma _{A}^{2}$ be given for simplicity, then
  \[\Sigma _{A}^{2}=\sigma _{A}^{2}+\rho \left( A,s \right),~~\Sigma _{B}^{2}=\sigma _{B}^{2}+\rho \left( B,s \right)\]
More specifically,
\[\Sigma _{A}^{2}=\sigma _{A}^{2}+\rho \left( A,s \right)=\sigma _{A}^{2}+\sigma _{u}^{2}-2\bar{u}\bar{A}+\overline{{{\left\{ A,u \right\}}_{ir}}}\]
As a consequence of above calculations, it leads to the complete result
  \[\Sigma _{A}^{2}\Sigma _{B}^{2}\equiv \sigma _{A}^{2}\sigma _{B}^{2}+\rho \left( A,s \right)\sigma _{B}^{2}+\sigma _{A}^{2}\rho \left( B,s \right)+\rho \left( A,s \right)\rho \left( B,s \right)\]
In another way,
\begin{align}\label{eq13}
  & \Delta {{{\hat{A}}}^{\left( s \right)}}\Delta {{{\hat{B}}}^{\left( s \right)}}=\frac{1}{2}{{\left[ \Delta {{\hat{A}}^{\left( s \right)}},\Delta {{\hat{B}}^{\left( s \right)}} \right]}_{QPB}}+\frac{1}{2}{{\left\{ \Delta {{\hat{A}}^{\left( s \right)}},\Delta {{\hat{B}}^{\left( s \right)}} \right\}}_{ir}}
\end{align}
More precisely, the details is given by
\begin{align}
  & \left( \Delta {{\hat{A}}^{\left( s \right)}} \right)\left( \Delta {{\hat{B}}^{\left( s \right)}} \right)=\left( \Delta \hat{A}+\Delta u \right)\left( \Delta \hat{B}+\Delta v \right) \notag\\
 & =\left( \widehat{A}-\overline{A}+u-\overline{u} \right)\left( \hat{B}-\overline{B}+v-\overline{v} \right)  \notag\\
 & =\left( \hat{A}-\overline{A} \right)\left( \hat{B}-\overline{B} \right)+\left( \hat{A}-\overline{A} \right)\left( v-\overline{v} \right)  \notag\\
 & \begin{matrix}
   {} & {}  \\
\end{matrix}+\left( u-\overline{u} \right)\left( \hat{B}-\overline{B} \right)+\left( u-\overline{u} \right)\left( v-\overline{v} \right)  \notag
\end{align}
where $\Delta v=v-\overline{v},~~v=\lambda\left( \hat{B},s\right)$. Hence, taking the modulus of means, the expectation in terms of the antisymmetric part of \eqref{eq13} follows
\begin{align}
  & \left( \psi ,\Delta {{\hat{A}}^{\left( s \right)}}\Delta {{\hat{B}}^{\left( s \right)}}\psi  \right)-\left( \psi ,\Delta {{\hat{B}}^{\left( s \right)}}\Delta {{\hat{A}}^{\left( s \right)}}\psi  \right) =\left( \psi ,{{\left[ \Delta {{\hat{A}}^{\left( s \right)}},\Delta {{\hat{B}}^{\left( s \right)}} \right]}_{QPB}}\psi  \right)  \notag\\
 & =\overline{AB}-\overline{BA}+\overline{B}\overline{A}-\overline{A}\overline{B}-\overline{A}\overline{v}+\overline{uB}-\overline{Bu}-\overline{u}\overline{B}+\overline{uv}-\overline{u}\overline{v} \notag\\
 & +\overline{B}\overline{u}+\overline{Av}-\overline{vA}+\overline{v}\overline{A}-\overline{vu}+\overline{u}\overline{v} \notag\\
 & =\overline{AB}-\overline{BA}+\overline{uB}-\overline{vA}+\overline{Av}-\overline{Bu} \notag\\
 & =\overline{AB-BA+uB-vA+Av-Bu}  \notag\\
 & =\overline{{{\left[\hat{A},\hat{B}  \right]}_{QPB}}+\left\langle \hat{A}:s:\hat{B} \right\rangle +{{\left[\hat{A},\hat{B}   \right]}_{QPB}}s}  \notag\\
 & =\overline{\left[\hat{A},\hat{B} \right]+2{{\left[ \hat{A},\hat{B}  \right]}_{QPB}}s}  \notag
\end{align}
where
 $\left\langle \hat{A}:s:\hat{B} \right\rangle =u\hat{B}-v\hat{A},~{{\left[ \hat{A},\hat{B}   \right]}_{QPB}}s=\hat{A}v-\hat{B}u$.
 Meanwhile, the symmetric part of \eqref{eq13} is evaluated as
\begin{align}
  & \left( \psi ,\Delta {{\hat{A}}^{\left( s \right)}}\Delta {{\hat{B}}^{\left( s \right)}}\psi  \right)+\left( \psi ,\Delta {{\hat{B}}^{\left( s \right)}}\Delta {{\hat{A}}^{\left( s \right)}}\psi  \right) =\left( \psi ,{{\left\{ \Delta {{\hat{A}}^{\left( s \right)}},\Delta {{\hat{B}}^{\left( s \right)}} \right\}}_{ir}}\psi  \right)  \notag\\
 & =\overline{AB}-\overline{A}\overline{B}+\overline{Av}-\overline{A}\overline{v}+\overline{uB}-\overline{u}\overline{B}+\overline{uv}-\overline{u}\overline{v} \notag\\
 & +\overline{BA}-\overline{B}\overline{A}+\overline{Bu}-\overline{B}\overline{u}+\overline{vA}-\overline{v}\overline{A}+\overline{vu}-\overline{u}\overline{v} \notag\\
 & =\overline{\left\{\hat{A},\hat{B}  \right\}+2uv}-2\left( \overline{A}\overline{v}+\overline{u}\overline{B}+\overline{u}\overline{v}+\overline{A}\overline{B} \right)  \notag\\
 &=\overline{\left\{ \hat{A},\hat{B} \right\}}+2\left( \overline{uv}-\left( \overline{A}\left( \overline{v}+\overline{B} \right)+\overline{u}\left( \overline{B}+\overline{v} \right) \right) \right)\notag
\end{align}
where
$\left( \hat{A}:s:\hat{B} \right)=v\hat{A}+u\hat{B},~~{{\left\{\hat{A},\hat{B} \right\}}_{ir}}s=\hat{A}v+\hat{B}u $.  Above all, summing up all the calculations, then
\begin{align}
  & \left( \psi ,\Delta {{\hat{A}}^{\left( s \right)}}\Delta {{\hat{B}}^{\left( s \right)}}\psi  \right)=\frac{1}{2}\left( \psi ,{{\left[ \Delta {{\hat{A}}^{\left( s \right)}},\Delta {{\hat{B}}^{\left( s \right)}} \right]}_{QPB}}\psi  \right)+\frac{1}{2}\left( \psi ,{{\left\{ \Delta {{\hat{A}}^{\left( s \right)}},\Delta {{\hat{B}}^{\left( s \right)}} \right\}}_{ir}}\psi  \right) \notag\\
 & =\left( \psi ,\frac{{{\left\{ \Delta {{\hat{A}}^{\left( s \right)}},\Delta {{\hat{B}}^{\left( s \right)}} \right\}}_{ir}}}{2}\psi  \right)+\sqrt{-1}\left( \psi ,\frac{{{\left[ \Delta {{\hat{A}}^{\left( s \right)}},\Delta {{\hat{B}}^{\left( s \right)}} \right]}_{QPB}}}{2\sqrt{-1}}\psi  \right)\notag \\
 & =\left( \frac{1}{2}\overline{\left\{\hat{A},\hat{B} \right\}}+\overline{uv}-\left( \overline{A}\left( \overline{v}+\overline{B} \right)+\overline{u}\left( \overline{B}+\overline{v} \right) \right) \right) \notag\\
 & \begin{matrix}
   {} & {} & {} & {}  \\
\end{matrix}+\sqrt{-1}\left( \frac{\overline{\left[ \hat{A},\hat{B}  \right]}+2\overline{{{\left[ \hat{A},\hat{B}  \right]}_{QPB}}s}}{2\sqrt{-1}} \right) \notag\\
 & =\left( \frac{1}{2}\overline{\left\{\hat{A},\hat{B} \right\}}+\theta \left( s,\hat{A},\hat{B} \right) \right)+\sqrt{-1}\left( \frac{\overline{\left[\hat{A},\hat{B}  \right]}}{2\sqrt{-1}}+\vartheta \left( s,\hat{A},\hat{B} \right) \right) \notag
\end{align}
where two notations are introduced to simplify the formula above
\begin{align}
  & \theta \left( s,\hat{A},\hat{B}\right)=\overline{uv}-\left( \left( \overline{A}+\overline{u} \right)\left( \overline{B}+\overline{v} \right) \right) \notag\\
 & \vartheta \left( s,\hat{A},\hat{B} \right)=-\sqrt{-1}\overline{{{\left[ \hat{A},\hat{B} \right]}_{QPB}}s}\notag
\end{align}
Taking the square modulus of it leads to the result,
\[{{\left| \left( \psi ,\Delta {{\hat{A}}^{\left( s \right)}}\Delta {{\hat{B}}^{\left( s \right)}}\psi  \right) \right|}^{2}}={{\left| \left( \frac{1}{2}\overline{\left\{ \hat{A},\hat{B}  \right\}}+\theta \left( s,\hat{A},\hat{B} \right) \right) \right|}^{2}}+{{\left| \left( \frac{\overline{\left[\hat{A},\hat{B} \right]}}{2\sqrt{-1}}+\vartheta \left( s,\hat{A},\hat{B}\right) \right) \right|}^{2}}\]
We let
\begin{align}
  & \Theta \left( s,\hat{A},\hat{B} \right)=\rho \left( A,s \right)\sigma _{B}^{2}+\sigma _{A}^{2}\rho \left( B,s \right)+\rho \left( A,s \right)\rho \left( B,s \right) \notag \\
 & \Xi \left( s,\hat{A},\hat{B} \right)={{\left| \left( \frac{1}{2}\overline{\left\{ \hat{A},\hat{B}  \right\}}+\theta \left( s,\hat{A},\hat{B} \right) \right) \right|}^{2}}+{{\left| \left( \frac{\overline{\left[\hat{A},\hat{B} \right]}}{2\sqrt{-1}}+\vartheta \left( s,\hat{A},\hat{B} \right) \right) \right|}^{2}} \notag
\end{align}be denoted, with the geometric Cauchy-Schwarz inequality
\[\Sigma _{A}^{2}\Sigma _{B}^{2}\equiv {{\left( \Delta {{A}^{\left( s \right)}} \right)}^{2}}{{\left( \Delta {{B}^{\left( s \right)}} \right)}^{2}}=\overline{{{\left( \Delta {{\hat{A}}^{\left( s \right)}} \right)}^{2}}}~\overline{{{\left( \Delta {{\hat{B}}^{\left( s \right)}} \right)}^{2}}}\ge {{\left| \overline{\Delta {{\hat{A}}^{\left( s \right)}}\Delta {{\hat{B}}^{\left( s \right)}}} \right|}^{2}}\]
such  that
$\Sigma _{A}^{2}\Sigma _{B}^{2}\equiv \sigma _{A}^{2}\sigma _{B}^{2}+\Theta \left( s,\hat{A},\hat{B} \right)=\Xi \left( s,\hat{A},\hat{B}\right)+\epsilon\left( s,\hat{A},\hat{B} \right)$, where $\epsilon\left( s,\hat{A},\hat{B} \right)\geq0$.  Therefore, we conclude that the equality holds in a state
\begin{align}
 \Sigma _{A}^{2}\Sigma _{B}^{2} & \equiv \sigma _{A}^{2}\sigma _{B}^{2}+\rho \left( A,s \right)\sigma _{B}^{2}+\sigma _{A}^{2}\rho \left( B,s \right)+\rho \left( A,s \right)\rho \left( B,s \right) \notag\\
 & ={{\left| \left( \frac{1}{2}\overline{\left\{\hat{A},\hat{B} \right\}}+\theta \left( s,\hat{A},\hat{B}\right) \right) \right|}^{2}}+{{\left| \left( \frac{\overline{\left[ \hat{A},\hat{B} \right]}}{2\sqrt{-1}}+\vartheta \left( s,\hat{A},\hat{B}\right) \right) \right|}^{2}}+\epsilon\left( s,\hat{A},\hat{B} \right)\notag
\end{align}
Thus, we complete the proof.
\end{proof}
\end{theorem}
Obviously, there comes a lot of discussions based on the theorem \ref{t5}, in particular, the analysis on the position operator and momentum operator, for this aim, we have to take several steps to achieve this purpose. According to the theorem \ref{t5} for $\epsilon\left( s,\hat{A},\hat{B} \right)>0$ in general, it clearly shows that
\[\sigma _{A}^{2}\sigma _{B}^{2}+\Theta \left( s,\hat{A},\hat{B} \right)\ge {{\left| \left( \frac{\overline{\left[\hat{A},\hat{B} \right]}}{2\sqrt{-1}}+\vartheta \left( s,\hat{A},\hat{B} \right) \right) \right|}^{2}}\]
And then the variance is
\[\wp \left( s,\hat{A},\hat{B} \right)=\Xi \left( s,\hat{A},\hat{B} \right) -\Theta \left( s,\hat{A},\hat{B}\right)+\epsilon\left( s,\hat{A},\hat{B} \right)\]Due to the join of the environment variable $s$, for a real quantum reality, supposed that
$\epsilon\left( s,\hat{A},\hat{B} \right)=0$ as a critical quantum condition for quantum operators $\hat{A},\hat{B} $, then it leads to the equation
\begin{align}
&  \sigma _{A}^{2}\sigma _{B}^{2}+\rho \left( A,s \right)\sigma _{B}^{2}+\sigma _{A}^{2}\rho \left( B,s \right)+\rho \left( A,s \right)\rho \left( B,s \right) \notag\\
 & ={{\left| \left( \frac{1}{2}\overline{\left\{\hat{A},\hat{B} \right\}}+\theta \left( s,\hat{A},\hat{B}\right) \right) \right|}^{2}}+{{\left| \left( \frac{\overline{\left[ \hat{A},\hat{B} \right]}}{2\sqrt{-1}}+\vartheta \left( s,\hat{A},\hat{B}\right) \right) \right|}^{2}}\notag
\end{align}
Such condition is valid to consider, in this case, if $\rho \left( A,s \right)=0$ is given, then it can be simplified as
$$\sigma _{A}^{2}\left(\sigma _{B}^{2}+\rho \left( B,s \right)\right)
 ={{\left| \left( \frac{1}{2}\overline{\left\{\hat{A},\hat{B} \right\}}+\theta \left( s,\hat{A},\hat{B}\right) \right) \right|}^{2}}+{{\left| \left( \frac{\overline{\left[ \hat{A},\hat{B} \right]}}{2\sqrt{-1}}+\vartheta \left( s,\hat{A},\hat{B}\right) \right) \right|}^{2}}$$
In this simple case, furthermore, if $u$ vanishes, then $\theta \left( s,\hat{A},\hat{B}\right)=-\overline{A} \left( \overline{B}+\overline{v} \right)$, and $\vartheta \left( s,\hat{A},\hat{B} \right)=-\sqrt{-1}\overline{{{\left[ \hat{A},\hat{B} \right]}_{QPB}}s}$. As a result of this,
it is clear to see that
$$\sigma _{A}^{2}\left(\sigma _{B}^{2}+\rho \left( B,s \right)\right)
 \geq{{\left| \left( \frac{\overline{\left[ \hat{A},\hat{B} \right]}}{2\sqrt{-1}}+\vartheta \left( s,\hat{A},\hat{B}\right) \right) \right|}^{2}}$$

First of all, we take the position operator and momentum operator into account, therefore, let $\hat{x}=x,\hat{p}=-\sqrt{-1}\hbar \frac{d}{dx}$ together be given, then
$\Theta \left( s,\hat{A},\hat{B} \right)=\rho \left( A,s \right)\sigma _{B}^{2}+\sigma _{A}^{2}\rho \left( B,s \right)+\rho \left( A,s \right)\rho \left( B,s \right)$  becomes
$$\Theta \left( s,x,\hat{p} \right)=\rho \left( x,s \right)\sigma _{\hat{p}}^{2}+\sigma _{x}^{2}\rho \left( \hat{p},s \right)+\rho \left( x,s \right)\rho \left( \hat{p},s \right)$$
where ${{\left[ x,s \right]}_{QPB}}=0$, then $\sigma _{u}^{2}-2\bar{u}\overline{x}+\overline{{{\left\{ x,u \right\}}_{ir}}}=0$, hence,
it marks $\rho \left( x,s \right)=0$ here, $\Theta \left( s,x,\hat{p} \right)=\sigma _{x}^{2}\rho \left( \hat{p},s \right)$, therefore,
$\rho \left( \hat{p},s \right)=\sigma _{u}^{2}-2\bar{u}\overline{\hat{p}}+\overline{{{\left\{ \hat{p},u \right\}}_{ir}}}$, then it gets
$$\Theta \left( s,x,\hat{p} \right)=\sigma _{x}^{2}\rho \left( \hat{p},s \right)=\sigma _{x}^{2}\sigma _{u}^{2}-2\sigma _{x}^{2}\bar{u}\overline{\hat{p}}+\sigma _{x}^{2}\overline{{{\left\{ \hat{p},u \right\}}_{ir}}}$$
In another aspects, it has
$$\Xi \left( s,x,\hat{p} \right)={{\left| \left( \frac{1}{2}\overline{\left\{ x,\hat{p} \right\}}+\theta \left( s,x,\hat{p} \right) \right) \right|}^{2}}+{{\left| \left( \frac{\overline{\left[ x,\hat{p} \right]}}{2\sqrt{-1}}+\vartheta \left( s,x,\hat{p} \right) \right) \right|}^{2}}$$where

 $\theta \left( s,x,\hat{p} \right)=\overline{uv}-\left( \left( \overline{x}+\bar{u} \right)\left( \overline{\hat{p}}+\bar{v} \right) \right) $

$\vartheta \left( s,x,\hat{p} \right)=-\sqrt{-1}\overline{{{\left[ x,\hat{p} \right]}_{QPB}}s}=-\sqrt{-1}\overline{{{\left[ x,\hat{p} \right]}_{QPB}}s}=\hbar \overline{s}$

\subsection{the case of the mean vanishes}

As we did previously, we consider the mean vanishes,
$\overline{X}$ and $\hat{X}=\hat{A},\hat{B}$, and then
$\sigma _{A}^{2}=\overline{{{A}^{2}}}$, others go as the same, then
\[\Sigma _{A}^{2}=\sigma _{A}^{2}+\rho \left( A,s \right)=\sigma _{A}^{2}+\sigma _{u}^{2}+\overline{{{\left\{ A,u \right\}}_{ir}}}\]where entanglement term becomes $\rho \left( A,s \right)=\sigma _{u}^{2}+\overline{{{\left\{ A,u \right\}}_{ir}}}$ in a simple form, it goes to the operator $\hat{B}$,
\[\Sigma _{B}^{2}=\sigma _{B}^{2}+\rho \left( B,s \right)=\sigma _{B}^{2}+\sigma _{v}^{2}+\overline{{{\left\{ B,v \right\}}_{ir}}}\]and
$\theta \left( s,\hat{A},\hat{B}\right)=\overline{uv}-\overline{u}~\overline{v}$,
If $u,v$ perform mutual independence, then $\theta \left( s,\hat{A},\hat{B} \right)=0$,
and it makes \[\Xi \left( s,\hat{A},\hat{B} \right)={{\left| \frac{1}{2}\overline{\left\{ \hat{A},\hat{B} \right\}} \right|}^{2}}+{{\left| \left( \frac{\overline{\left[  \hat{A},\hat{B}\right]}}{2\sqrt{-1}}+\vartheta \left( s,\hat{A},\hat{B} \right) \right) \right|}^{2}}\]a simple form, thus, on this case, plugging all parts into the theorem \ref{t5}, we obtain the QGR on the case of the mean vanishes.

Obviously, seen from the theorem \ref{t5}, the entanglement term does exists between the environment and the observables represented by entanglement function
\[\rho \left( X,s \right)=\sigma _{u}^{2}-2\overline{u}\overline{X}+\overline{{{\left\{\hat{X},u \right\}}_{ir}}}\]
where $\hat{X}$ can be arbitrarily chosen as a Hermitian operators, and  $u=\lambda\left( \hat{X},s\right)$. The theorem \ref{t5} at $\epsilon\left( s,\hat{A},\hat{B} \right)=0$ truly indicates that a fact for certainty with measurement
\[\sigma _{A}^{2}\sigma _{B}^{2}=\Xi \left( s,\hat{A},\hat{B} \right)-\Theta \left( s,\hat{A},\hat{B} \right)\]always holds in quantum mechanics, and $\Xi \left( s,\hat{A},\hat{B}\right)>\Theta \left( s,\hat{A},\hat{B} \right)$.

The theorem \ref{t5} indicates that the generalized variance has a deep connection with the environment variable
\[\Sigma _{X}^{2}=\sigma _{X}^{2}+\rho \left( X,s \right)=\sigma _{X}^{2}+\sigma _{u}^{2}-2\bar{u}\overline{X}+\overline{{{\left\{ \hat{X},u \right\}}_{ir}}}\]
It reveals that the environment variable occupies an important place in the quantum mechanics.

It emphasizes that the quantum geometric bracket pictures the intrinsic interaction between the observables and the environment. With this perspective, it certainly leads to the consequence below.

{\bf{Geomertainty principle:}}
The operators corresponding to certain observables satisfy the quantum geomertainty relation, then there admits the outcome of each measurement with more certain.

According to the QGR, we have an interpretation for the measurement with certainty.
\[\Delta f=\varpi \left( s,\hat{f},\hat{g} \right)/\Delta g\]
 The QGR definitely shows that they must all have determinate values.
The operators obey the covariant evolution and the geomertainty principle, then it forms a complete description of determinism of observables.

Assume that if for any Hermitian operator ${\hat {A}}$ that is not an angle variable and ${\hat {B}}$ is not the derivative with respect to this variable. Then the Robertson uncertainty relation is generally formulated as the relation
$$\sigma(A, \psi) \sigma(B, \psi) \geq \frac{|\langle\psi|[\hat{A},\hat{B}]_{QPB}| \psi\rangle|}{2}$$
for any observables $\hat{A},\hat{B}$ and any state $\psi$. As our further result shows, it has deeply interpretation given by
\[{{\sigma }_{A}}{{\sigma }_{B}}\ge \sqrt{{\Gamma _{ne}^{2}\left( s,\hat{A},\hat{B} \right)}-\Theta\left( s,\hat{A},\hat{B}\right)}\]
where ${{\Gamma }_{ne}}\left( s,\hat{A},\hat{B} \right)=\frac{\left\langle \left[ \hat{A},\hat{B} \right] \right\rangle }{2\sqrt{-1}}+\vartheta \left( s,\hat{A},\hat{B} \right)$,  it implies that the Robertson uncertainty relation is a very narrow explanation.

As Heisenberg emphasized, the non-commutativity implies the uncertainty principle, Heisenberg showed that the QPB defined by commutation relation implies an uncertainty. Now, the QPB has been completely replaced by the QCPB, it leads to a further complete theory.  Modern studies generally agree that the uncertainty is caused by the environment.  Due to the QCPB as a further development of the QPB, as a result of the environment variable $s$ emerges naturally, hence,  we would think of a complete expression for such intentions to explain exactly how this happens in quantum mechanics. Up to now, we will see that the QCPB  defined by generalized geometric commutation relation as an application provided a clear physical covariant interpretation for such uncertainty and its origin in quantum methods.

 Similarly, the same procedure goes for Robertson uncertainty relation in terms of $x$ and $\hat{p}$
\[\vartheta \left( x \right)\vartheta \left( p \right)=\sqrt{\Xi \left( s,\hat{x},\hat{p}\right)-\Theta\left( s,\hat{x},\hat{p} \right)+\epsilon\left( s,\hat{x},\hat{p} \right)}\]
where $\vartheta $ can be $\Delta $ or $\sigma $.
We positively answer that the existence of indeterminacy for some
measurements are rightly caused by the environment that can be certainly expressed in a quantitative form by quantum geomertainty relation.
If $\hat{f}=x,~\hat{g}=\hat{p}$ are taken, then it states that the more precisely the position of some particle is determined, the more precisely its momentum can be known, but the environment variable $s$ is the key to all such measurements.

Clearly, we can verify that there always exists geometric variables to derive the uncertainty relation which should be replaced by a quantum identical equation. Quantum geomertainty relation as a quantum geometric certainty equality can well explain the outcome of each measurement with certainty. In this way, we eliminate some ambiguities in formulation.

From above derivations, it evidently implies that the measurement with uncertainty directly comes from the quantum geometric bracket which mainly describes the interaction between the observables and the environment variable or just environment.

\subsection{The entanglement term and the curvature}

Obviously, seen from the theorem \ref{t5}, the entanglement term does exist between the environment and the observables represented by entanglement function
\begin{equation}\label{eq17}
  \rho \left( X,s \right)=\sigma _{u}^{2}-2\overline{u}\overline{X}+\overline{{{\left\{\widehat{X},u \right\}}_{ir}}}
\end{equation}
where $\hat{X}$ can be arbitrarily chosen as a Hermitian operators, and  $u=\lambda\left( \hat{X},s\right)$, $\sigma _{u}^{2}=\overline{{{u}^{2}}}-{{\bar{u}}^{2}}$.
Let's take the geomentum operator $\hat{X}=\hat{p}=-\sqrt{-1}\hbar D$ into consideration, to consider its components,  ${{\hat{X}}_{j}}={{\hat{p}}_{j}}=-\sqrt{-1}\hbar {{D}_{j}}$,
then
${{u}_{j}}={{\hat{p}}_{j}}s=-\sqrt{-1}\hbar {{D}_{j}}s$.
and $u=\hat{p}s=-\sqrt{-1}\hbar Ds$, hence, by a directly computation, we get
\[{{u}^{2}}={{\left( \hat{p}s \right)}^{2}}=-{{\hbar }^{2}}{{\left| Ds \right|}^{2}},~{{\bar{u}}^{2}}=-{{\hbar }^{2}}{{\overline{Ds}}^{2}}\]
Therefore, the deviation in terms of $u$ is \[\sigma _{u}^{2}=\overline{{{u}^{2}}}-{{\bar{u}}^{2}}={{\hbar }^{2}}{{\overline{Ds}}^{2}}-{{\hbar }^{2}}\overline{{{\left| Ds \right|}^{2}}}={{\hbar }^{2}}\left( {{\overline{Ds}}^{2}}-\overline{{{\left| Ds \right|}^{2}}} \right)\]
And then, it also has a series of calculations given by
$\bar{u}\bar{X}=-{{\hbar }^{2}}\overline{Ds}\overline{D}$,
and
\begin{align}
 {{\left\{ X,u \right\}}_{ir}} &=Xu+uX=-\sqrt{-1}\hbar Du-u\sqrt{-1}\hbar D \notag\\
 & =-{{\hbar }^{2}}\left( {{D}^{2}}s+DsD \right) \notag
\end{align}Thusly, it yields
\[\overline{{{\left\{ X,u \right\}}_{ir}}}=-{{\hbar }^{2}}\left( \overline{{{D}^{2}}s}+\overline{DsD} \right)\]
Above all, the entanglement function in terms of the geomentum operator writes
\begin{align}\label{eq27}
 \rho \left( p,s \right) &=\sigma _{u}^{2}-2\bar{u}\bar{p}+\overline{{{\left\{\hat{p},u \right\}}_{ir}}} \\
 & ={{\hbar }^{2}}\left( {{\overline{Ds}}^{2}}-\overline{{{\left| Ds \right|}^{2}}} \right)+2{{\hbar }^{2}}\overline{Ds}\overline{D}-{{\hbar }^{2}}\left( \overline{{{D}^{2}}s}+\overline{DsD} \right) \notag\\
 & ={{\hbar }^{2}}\left( {{\overline{Ds}}^{2}}-\overline{{{\left| Ds \right|}^{2}}}+2\overline{Ds}\overline{D}-\overline{{{D}^{2}}s}-\overline{DsD} \right) \notag
\end{align}
As a matter of fact,
\[\rho \left( p,s \right)/{{\hbar }^{2}}={{\overline{Ds}}^{2}}-\overline{{{\left| Ds \right|}^{2}}}+2\overline{Ds}\overline{D}-\overline{{{D}^{2}}s}-\overline{DsD}\]
can be realized as the curvature terms created by a manifold or space represented by this environment variable $s$.

Let's consider position operator in one dimensional, namely, $X=x$, then ${{\left[ s,x \right]}_{QPB}}=0$, it leads to
$\rho \left( x,s \right)=0$.
Note that the geomentum operator and the structure function or environment variable are truly existed, then it implies entanglement function \eqref{eq17} is a real existence for quantum mechanics.  In particular, the entanglement function in terms of the geomentum operator tells us that the measurement with certainty relates the space curvature.

\section{Universally formulation of QGR for position and momentum }
The uncertainty principle is any of a variety of mathematical inequalities asserting a fundamental limit to the precision with which certain pairs of physical properties of a particle, known as complementary variables or canonically conjugate variables such as position $x$ and momentum $p$, can be known or, depending on interpretation, to what extent such conjugate properties maintain their approximate meaning, as the mathematical framework of quantum physics does not support the notion of simultaneously well-defined conjugate properties expressed by a single value.

Universally formulation of quantum geomertainty relation as a quantum geometric certainty relation can be held under the framework of QCPB.  In the guidance of the QCPB, the Robertson uncertainty relation \cite{27} is rewritten in a certainty equality form generally formulated as the relation
\begin{align}
\sigma \left( \hat{A},\psi  \right)\sigma \left(\hat{B},\psi  \right)=\varpi \left( s,\hat{A},\hat{B} \right)=\sqrt{\wp \left( s,\hat{A},\hat{B} \right)}\notag
\end{align}
for any observables $\hat{A}, \hat{B}$ and any state $\psi$, where the standard deviation $\sigma(\hat{X},\psi)$ of an observable $\hat{X}$ in state $\psi$ is
defined by
\[{{\sigma }^{2}}\left(\hat{X},\psi  \right)=\left\langle  \psi  \right|{{\hat{X}}^{2}}\left| \psi  \right\rangle -\left\langle  \psi  \right|\hat{X}{{\left| \psi  \right\rangle }^{2}}\]

As an application of the QGR in terms of the position and momentum, we mainly concentrate on the one-dimensional case.
In the beginning, we give the QGR for position and momentum in a state as follows.

{\bf{Quantum geomertainty relation in terms of position-momentum:}}
 The QGR in terms of quantum operator ${x},~\hat{p}$ is
  \begin{equation}\label{eq5}
    \Delta x\cdot\Delta p=\varpi \left( s,{x},\hat{p} \right)=\sqrt{\wp \left( s,{x},\hat{p} \right)}\
  \end{equation}
where $\wp \left( s,{x},\hat{p} \right)=\Xi \left( s,{x},\hat{p}\right)-\Theta \left( s,{x},\hat{p} \right)+\epsilon \left( s,{x},\hat{p} \right)$.

Transparently, as we can see from the certainty formula,  the quantum geomertainty relation allows one to predict the position and momentum with certainty.

As we calculated in the \eqref{eq27}, the entanglement function in terms of the geomentum operator states that the curvature of the spacetime included in the $\Theta \left( s,{x},\hat{p} \right)$ is inevitable to produce an effect on the quantum measurement. Actually, the QGR for position and momentum in a state has a complex explanation according to the curvature fact.   We can conclude that the entanglement function \eqref{eq17} in terms of the geomentum operator \[\rho \left( p,s \right)=\sigma _{u}^{2}-2\bar{u}\overline{p}+\overline{{{\left\{ \hat{p},u \right\}}_{ir}}}\] rightly represents the curvature of the spacetime, where  $u=\hat{p}s$.

\subsection{One-dimensional case}
In this section, we mainly discuss the one-dimensional case, in this case, we use $i^{2}=-1$ to simplify the equations.
Geometric canonical commutation relation \cite{15} will be used for this QGR shown as \[\left[ \hat{{{x}_{i}}},\hat{{{p}_{j}}} \right]= i\hbar{{D}_{j}}{{x}_{i}}=i\hbar \left( {{\delta }_{ij}}+{{x}_{i}}A_{j}\right) \]
One-dimensional case is given by
$\left[ x,\hat{p} \right]=i\hbar b\left( x \right)$, where $s$ is a geometric potential function and $b\left( x \right)=1+x\mu$, and $\mu=\frac{ds}{dx}$.

As we know,  the classical result is calculated by the QPB
$$\Delta x\cdot \Delta p\ge \left| \frac{\overline{{{\left[ x,\hat{{{p}}} \right]}_{QPB}}}}{2i} \right|=\frac{\hbar }{2}$$
By using $\left[ x,\hat{p} \right]=i\hbar \left( 1+x\mu\right)$, we have
\begin{align}
\left| \frac{\overline{\left[ x,\hat{p} \right]}}{2i} \right|& =\left| \frac{\overline{{{\left[ x,\hat{p} \right]}_{QPB}}}+\overline{G\left( s,x,\hat{p} \right)}}{2i} \right|=\frac{\hbar }{2}\left| \overline{ 1+x\mu} \right|=\frac{\hbar }{2}\left| 1+\overline{x\mu} \right| \notag
\end{align}

\subsubsection{In geomentum operator}
When we apply above theorem to the position operator and momentum operator, it gives the specific results.
Subsequently, let's compute the geometric anti-commutator (GAC) in terms of
$\hat{x},~\hat{p}$, that is, $\left\{ x,\hat{p}  \right\}={{\left\{ x,\hat{p} \right\}}_{ir}}+Z\left( s,x,\hat{p}  \right)$, as a consequence, taking the geomentum operator $\hat{p} =-i\hbar \frac{\text{D}}{dx}$ into account, we obtain the calculation of the classical anti-commutator
\begin{align}
  {{\left\{ x,\hat{p}  \right\}}_{ir}}\psi & =\left( x\hat{p} +\hat{p} x \right)\psi =-i\hbar \left( x\frac{\text{D}}{dx}\psi +\frac{\text{D}}{dx}\left( x\psi  \right) \right) \notag\\
 & =-2i\hbar \left( x\frac{d}{dx}+x\mu+1/2 \right)\psi \notag
\end{align}Then it gets
${{\left\{ x,\hat{p} \right\}}_{ir}}=-2i\hbar \left( x\frac{d}{dx}+1/2+x\mu \right)$ and anti-geomutator in terms of $x,\hat{p}$ is
\begin{align}
 Z\left( s,x,\hat{p}  \right)\psi &=x{{\left\{ s,\hat{p} \right\}}_{ir}}\psi +\hat{p} {{\left\{ s,x \right\}}_{ir}}\psi=x\left( s\hat{p} +\hat{p} s \right)\psi +2\hat{p} \left( xs\psi  \right) \notag\\
 & =-i\hbar \left( 4xs\frac{d}{dx}+4xs\mu+3x\mu+2s \right)\psi  \notag
\end{align}
Hence, the GAC is given by
\[\left\{ x,\hat{p} \right\}=-i\hbar \left( 2x\frac{d}{dx}+1+K\left( s,x \right) \right)\]
where
$K\left( s,x \right)=x\left( 2\frac{d{{s}^{2}}}{dx}+4s\frac{d}{dx}+5\mu \right)+2s$.

Based on the quantum geomertainty relation in terms of position-momentum and the theorem \ref{t5}, we have a series of calculations given by
\begin{align}
  & \Xi \left( s,x,p \right)={{\left( \left\langle \left\{ {x},\hat{p} \right\} \right\rangle /2+\theta \left( s,x,p \right) \right)}^{2}}+{{\left( \left\langle \left[ {x},\hat{p} \right] \right\rangle /2i+\vartheta \left( s,x,p \right) \right)}^{2}}  \notag
\end{align}
where

$\theta \left( s,\hat{x},\hat{p}\right)=- \overline{x}\left( \overline{p}+\overline{ps} \right)   $,

$\vartheta \left( s,\hat{x},\hat{p}\right)=-i\overline{{{\left[ x,p \right]}_{QPB}}s}=\hbar \overline{s}$,

$\Theta \left( s,\hat{x},\hat{p} \right)=\sigma _{x}^{2}\rho \left( s,p \right)$, \\
and the entanglement function in terms of the operator $\hat{X}$ is
\[\rho \left( X,s \right)=\sigma _{u}^{2}-2\bar{u}\bar{X}+\overline{{{\left\{ X,u \right\}}_{ir}}}\]for $X=x,\hat{p}$, and  $\rho \left(x,s \right)=0$.
Above all, plugging the calculations above into the quantum geomertainty relation in terms of position-momentum and the theorem \ref{t5}, we get the complete expression.
\begin{align}
  \Xi \left( s,\hat{x},\hat{p}\right)& ={{\left| \left( \frac{1}{2}\overline{\left\{ x,\hat{p} \right\}}+\theta \left( s,\hat{x},\hat{p} \right) \right) \right|}^{2}}+{{\left| \left( \frac{\overline{\left[ x,\hat{p} \right]}}{2i}+\vartheta \left( s,\hat{x},\hat{p} \right) \right) \right|}^{2}}  \notag\\
 & ={{\left| \left( -i\hbar \frac{1}{2}\overline{\left( 2x\frac{d}{dx}+K\left( s,x \right) \right)}-i\hbar /2+\theta \left( s,\hat{x},\hat{p} \right) \right) \right|}^{2}}+{{\left| \hbar \left( \overline{b\left( x \right)}/2+\overline{s} \right) \right|}^{2}}  \notag
\end{align}

\subsubsection{In classical momentum operator}
Actually, we can calculate above the geometric anti-commutator (GAC) in terms of
$\hat{x},~\hat{p}$ in classical momentum operator, let's see how different both they are. In this way, if taking $\hat{p}=-i\hbar \frac{d}{dx}$ into account,
\begin{align}
  {{\left\{ x,\hat{p} \right\}}_{ir}}\psi &=\left( x\hat{p}+\hat{p}x \right)\psi  =-2i\hbar \left( \frac{1}{2}+x\frac{d}{dx} \right)\psi  \notag
\end{align}
Then it leads to
${{\left\{ x,\hat{p} \right\}}_{ir}}=-2i\hbar \left( x\frac{d}{dx} +\frac{1}{2}\right)$.
Accordingly, the anti-geomutator can be calculated as
\begin{align}
 Z\left( s,x,\hat{p} \right)\psi &  =x\left( s\hat{p}+\hat{p}s \right)\psi +2\hat{p}\left( xs\psi  \right) \notag\\
 & =-xi\hbar \left( 4s\frac{d}{dx}+3q \right)\psi -2i\hbar s\psi  \notag
\end{align}
where

${{\left\{ s,\hat{p} \right\}}_{ir}}\psi =-i\hbar \left( 2s\frac{d}{dx}+q \right)\psi $,

$2\hat{p}\left( xs\psi  \right)=-2i\hbar \left( s+xq+xs\frac{d}{dx} \right)\psi  $.\\
Thusly, it gets
$Z\left( s,x,\hat{p} \right)=-xi\hbar \left( 4s\frac{d}{dx}+3q \right)-2i\hbar s$,
hence, the GAC is given by
$\left\{ x,\hat{p} \right\} =-2i\hbar \left( \frac{1}{2}+x\frac{d}{dx}+K\left( s,x \right) \right)$, where $K\left( s,x \right)=s+2sxd/dx+3xq/2$.

\section{ Concluding remarks}
In this paper, we have proposed quantum geomertainty relation (QGR) in deformation quantization formalism strongly relied on the QCPB proposed in \cite{15}. To achieve this, first it was necessary to study a general theory of the uncertainty relation and its developments based on the uncertainty relation which is strictly given by the QPB or the commutation.

As described in \cite{15}, the framework of QCPB is a complete theory as the complete description of physical reality that is able to predict experimental results with localized data, the environment variable in the quantum geometric bracket that exists naturally can be a chosen variables in the complete description of quantum mechanics. As a result of the quantum geomertainty relation (QGR), it pictures the much more complete description of physical reality.  To apply the quantum geomertainty relation (QGR) to the case of the position and momentum gives some useful results. This allowed us to formulate several certainty relations. Of course, further investigations in this direction are needed.
It is expected that the new results of the quantum geomertainty relation (QGR) will give a better understanding of the relations between quantum measurements of quantum mechanics.


\begin{thebibliography}{99}

\bibitem{1}
Heisenberg W 1927 Z. Phys. 43 172.  [English
translation by J. A. Wheeler and W. H. Zurek,
in Quantum Theory and Measurement, edited by
J. A. Wheeler and W. H. Zurek (Princeton UP,
Princeton, NJ, 1983), pp. 62-84].


Werner Heisenberg, The Physical Principles of the Quantum Theory, p. 20

\bibitem{2}
 Kennard E H 1927 Z. Phys. 44 326

\bibitem{3} Weyl H 1928 Gruppentheorie und Quantenmechanik (Leipzig: Hirzel Verlag)

\bibitem{4} Robertson H P 1929 Phys. Rev. 34 163

\bibitem{5} Robertson H P 1930 Phys. Rev. 35 667A

\bibitem{6} Schr\"{o}dinger E 1930 Sitz. Preus. Acad. Wiss. (Phys.-Math. Klasse) 296

\bibitem{7}Robertson H P 1934 Phys. Rev. 46 794

\bibitem{8}
 Trifonov D A 1999 The uncertainty way of generalization of coherent states Preprint
quant-ph/9912084 v5

\bibitem{9}  Trifonov D A 2001 Generalizations of Heisenberg uncertainty relation Preprint quant-ph/0112028

\bibitem{10} Bayen F, Flato M, Fronsdal M, Lichnerowicz A and Sternheimer D 1978 Ann. Phys.,
NY 111, 61
Bayen F, Flato M, Fronsdal M, Lichnerowicz A and Sternheimer D 1978 Ann. Phys., NY 111, 111
\bibitem{11} Sternheimer D 1998 Deformation quantization: twenty years after Particles, Fields and Gravitation ed J Rembieli´ nski (Woodbury, New York: American Institute of Physics), pp 107-145
\bibitem{12} Dito G and Sternheimer D 2002 Deformation quantization: genesis, developments and metamorphoses Preprint math.QA/0201168

\bibitem{13} Antonsen F 1997 Phys. Rev. D 56 920

\bibitem{14} Curtright T, Fairlie D and Zachos C 1998 Phys. Rev. D 58 025002\\
Curtright T, Uematsu T and Zachos C 2001 J. Math. Phys. 42 2396

\bibitem{15}

G Wang. Generalized geometric commutator theory and quantum geometric bracket and its uses [J].arXiv:2001.08566v3



\bibitem{16}
Einstein, A.; Podolsky, B.; Rosen, N. (1935). "Can Quantum-Mechanical Description of Physical Reality Be Considered Complete?". Physical Review. 47 (10): 777-780.

\bibitem{17}
Davidson, E. R. (1965), "On Derivations of the Uncertainty Principle", J. Chem. Phys., 42 (4): 1461–1462,

\bibitem{18}
 Maccone, Lorenzo; Pati, Arun K. (31 December 2014). "Stronger Uncertainty Relations for All Incompatible Observables". Physical Review Letters. 113 (26): 260401.

\bibitem{19}
Rozema, L. A.; Darabi, A.; Mahler, D. H.; Hayat, A.; Soudagar, Y.; Steinberg, A. M. (2012). "Violation of Heisenberg's Measurement-Disturbance Relationship by Weak Measurements". Physical Review Letters. 109 (10): 100404.

\bibitem{20}
 L. I. Mandelshtam, I. E. Tamm, The uncertainty relation between energy and time in nonrelativistic quantum mechanics, 1945.

\bibitem{21} Fujikawa, Kazuo (2012). "Universally valid Heisenberg uncertainty relation". Physical Review A. 85 (6): 062117.


\bibitem{22}
Busch, P.; Lahti, P.; Werner, R. F. (2013). "Proof of Heisenberg's Error-Disturbance Relation". Physical Review Letters. 111 (16): 160405.

\bibitem{23}
Huang, Yichen (10 August 2012). "Variance-based uncertainty relations". Physical Review A. 86 (2): 024101.

\bibitem{24}
Hilgevoord, Jan (1996). "The uncertainty principle for energy and time" (PDF). American Journal of Physics. 64 (12): 1451–1456.

\bibitem{25}
Busch, P.; Lahti, P.; Werner, R. F. (2014). "Heisenberg uncertainty for qubit measurements". Physical Review A. 89 (1): 012129.

\bibitem{26}
Werner, R. F. The uncertainty relation for joint measurement of position and momentum. Quantum Inf. Comput. 4, 546-562 (2004).


\bibitem{27}
Ozawa, Masanao (2003), "Universally valid reformulation of the Heisenberg uncertainty principle on noise and disturbance in measurement", Physical Review A, 67 (4): 42105,

Ozawa, M. Quantum limits of measurements and uncertainty principle. pp 3-17 in Bendjaballah,
C. et al. (eds) Quantum Aspects of Optical Communications. (Springer, Berlin, 1991).


Ozawa, M. Physical content of the Heisenberg uncertainty relation: limitation and reformulation. Phys. Lett. A 318, 21-29 (2003).
\bibitem{28}


Vakili B , Sepangi H R . Generalized uncertainty principle in Bianchi type I quantum cosmology [J]. Physics Letters B, 2007, 651(2-3):79-83.
\bibitem{29}


Maggiore M . A generalized uncertainty principle in quantum gravity[J]. Physics Letters B, 1993, 304(1-2):65-69.

\bibitem{30}
Buoninfante L , Luciano G G , Petruzziello L . Generalized Uncertainty Principle and Corpuscular Gravity[J]. European Physical Journal C, 2019.

\bibitem{31}
Branciard, C. Error-tradeoff and error-disturbance relations for incompatible quantum measurements [J]. Proc Natl Acad Sci U S A, 110(17):6742-6747.

\bibitem{32}
Busch, Paul, Lahti, Pekka, Werner, Reinhard F. Proof of Heisenberg's Error-Disturbance Relation [J]. Physical Review Letters, 111(16):160405.


\bibitem{33}
Kaneda F , Baek S Y , Ozawa M , et al. Experimental Test of Error-Disturbance Uncertainty Relations by Weak Measurement [J]. Physical Review Letters, 2014, 112(2):020402.

\bibitem{34}
Braunstein, Samuel L.; Caves, Carlton M.; Milburn, G.J. (April 1996). "Generalized Uncertainty Relations: Theory, Examples, and Lorentz Invariance". Annals of Physics. 247 (1): 135–173.

\bibitem{35}
 Louisell, W. H. (1963), "Amplitude and phase uncertainty relations", Physics Letters, 7 (1): 60–61,

\bibitem{36}
Bialynicki-Birula, I.; Mycielski, J. (1975), "Uncertainty Relations for Information Entropy in Wave Mechanics", Communications in Mathematical Physics, 44 (2): 129–132,

\bibitem{37}
Folland, Gerald; Sitaram, Alladi (May 1997), "The Uncertainty Principle: A Mathematical Survey", Journal of Fourier Analysis and Applications, 3 (3): 207–238,


\bibitem{38}
Hedenmalm, H. (2012), "Heisenberg's uncertainty principle in the sense of Beurling", J. Anal. Math., 118 (2): 691–702,

\bibitem{39}

D, Sen. The Uncertainty relations in quantum mechanics. Current Science. 2014, 107 (2): 203-218.

\bibitem{40}
Ozawa, M. Physical content of the Heisenberg uncertainty relation: Limitation and reformulation. Phys. Lett. A. 2003, 318, 21-29.

\bibitem{41}
A P Lund and H M Wiseman. Measuring measurement-disturbance relationships with weak values,New Journal of Physics. 2010, 11pp, 93011.


\bibitem{42}
M. Przanowski, F.J.
Turrubiates Uncertainty Relations in Deformation Quantization
J. Phys. 2002, A 35, page 10643.


\bibitem{43}
 Wheeler, J. A.  Zurek, W. H. (eds.) Qunatum Theory and Measurement. (Princeton Univ. Press,1983).
\bibitem{44}
Haroche, S.  Raimond J.-M. Exploring the Quantum (Oxford Univ. Press, 2006).
\bibitem{45}
Arthurs, E.  Goodman, M. S. Quantum correlations: a generalized Heisenberg uncertainty
relation. Phys. Rev. Lett. 60, 2447–2449 (1988).
\bibitem{46}
Ishikawa, S. Uncertainty relations in simultaneous measurements for arbitrary observables. Rep.
Math. Phys. 29, 257-273 (1991).

\bibitem{47}
Bj\"{o}rk, G.; S\"{o}derholm, J.; Trifonov, A.; Tsegaye, T.; Karlsson, A. (1999). "Complementarity and the uncertainty relations". Physical Review. A60 (3): 1878.
\bibitem{48}
Lund, A. P. Wiseman, H. M. Measuring measurement-disturbance relationships with weak values. New J. Phys. 12, 093011 (2010).
\bibitem{49}
 A. F. Ali, S. Das, and E. C. Vagenas, Phys. Lett. B678, 497 (2009).
\bibitem{50} S. Das, E. C. Vagenas, and A. F. Ali, Phys. Lett. B690, 407 (2010).
\bibitem{51} A. F. Ali, S. Das, and E. C. Vagenas, Phys. Rev. D 84, 044013 (2011).




\end{thebibliography}
\end{document}